\documentclass[5p,a4paper]{elsarticle}
\journal{Ultramicroscopy}
\usepackage[english]{babel}

\usepackage{amsmath, amssymb}
\usepackage{graphicx}
\usepackage[colorinlistoftodos]{todonotes}

\usepackage{siunitx}
\usepackage{subcaption}
\usepackage[font={small,sl},labelfont=bf]{caption}
\usepackage{placeins}

\definecolor{C0}{HTML}{1f77b4}
\definecolor{C1}{HTML}{ff7f0e}
\definecolor{C2}{HTML}{2ca02c}
\definecolor{C3}{HTML}{d62728}
\definecolor{C4}{HTML}{9467bd}
\definecolor{C5}{HTML}{8c564b}
\definecolor{C6}{HTML}{e377c2}
\definecolor{C7}{HTML}{7f7f7f}
\definecolor{C8}{HTML}{bcbd22}
\definecolor{C9}{HTML}{17becf}

\usepackage{csquotes}
\usepackage[colorlinks=true, allcolors=blue]{hyperref}
\newcommand{\subf}[1]{\textbf{(#1)}}
\begin{document}
\begin{frontmatter}


\title{Quantitative analysis of spectroscopic Low Energy Electron Microscopy data: High-dynamic range imaging, drift correction and cluster analysis}

\author[1]{T.A.\ de Jong\corref{cor1}}
\ead{jongt@physics.leidenuniv.nl}
\author[2]{D.N.L.\ Kok}
\author[1]{A.J.H.\ van der Torren}
\author[1]{H.\ Schopmans}
\author[1,3]{R.M.\ Tromp}
\author[1]{S.J.\ van der Molen}
\author[1]{J.\ Jobst}
\address[1]{Huygens-Kamerlingh Onnes Laboratorium, Leiden Institute of Physics, Leiden University, Niels Bohrweg 2, P.O. Box 9504, NL-2300 RA Leiden, The Netherlands}
\address[2]{Mathematical Institute, Leiden University, Niels Bohrweg 1, 23332CA Leiden, The Netherlands}
\address[3]{IBM T. J. Watson Research Center, 1101 Kitchawan Road, P.O. Box 218, Yorktown Heights, New York 10598, USA}
\cortext[cor1]{Corresponding Author}

\begin{abstract}
For many complex materials systems, low-energy electron microscopy (LEEM) offers detailed insights into morphology and crystallography by  naturally combining real-space and reciprocal-space information. Its unique strength, however, is that all measurements can easily be performed energy-dependently. Consequently, one should treat LEEM measurements as multi-dimensional, spectroscopic datasets rather than as images to fully harvest this potential.
Here we describe a measurement and data analysis approach to obtain such quantitative spectroscopic LEEM datasets with high lateral resolution. 
The employed detector correction and adjustment techniques enable measurement of true reflectivity values over four orders of magnitudes of intensity. 
Moreover, we show a drift correction algorithm, tailored for LEEM datasets with inverting contrast, that yields sub-pixel accuracy without special computational demands. 
Finally, we apply dimension reduction techniques to summarize the key spectroscopic features of datasets with hundreds of images into two single images that can easily be presented and interpreted intuitively. We use cluster analysis to automatically identify different materials within the field of view and to calculate average spectra per material.
We demonstrate these methods by analyzing bright-field and dark-field datasets of few-layer graphene grown on silicon carbide and provide a high-performance Python implementation. 
\end{abstract}

\end{frontmatter}

\section{Introduction}
Low Energy Electron Microscope (LEEM) is a surface science technique where images are formed from reflected electrons of low kinetic energy---down to single electronvolts. This is achieved by decelerating the electrons before they reach the sample and projecting them onto a pixelated detector after interaction with the sample. 
LEEM has proven to be a versatile tool, due to its damage-free, real-time imaging capabilities and its combination of electron diffraction with spectroscopic, and real-space information. 
This enables more advanced LEEM-based techniques such as dark-field imaging, where electrons from a single diffracted beam are used to create a real-space image, revealing spatial information on the atomic lattice of the sample~\cite{Bauer1989,dejong2018intrinsic}.

Aside from usage as an imaging tool, LEEM is frequently used as a tool for quantitative analysis of physical properties of a wide range of materials. 
Multi-dimensional datasets can be created by recording LEEM images as a function of one or more parameters such as interaction energy $E_0$, angle of incidence or temperature~\cite{flege2014intensity, jobst2015nanoscale}. Using this, a wide range of properties can be studied, for example, layer interaction, electron bands~\cite{jobst2016quantifying}, layer stacking~\cite{dejong2018intrinsic}, catalysis~\cite{Schmid1995}, plasmons~\cite{Frank2017}, and surface corrugation~\cite{locatelli2010corrugation}.

\begin{figure}[!hb]
\centering
\includegraphics[width=\columnwidth]{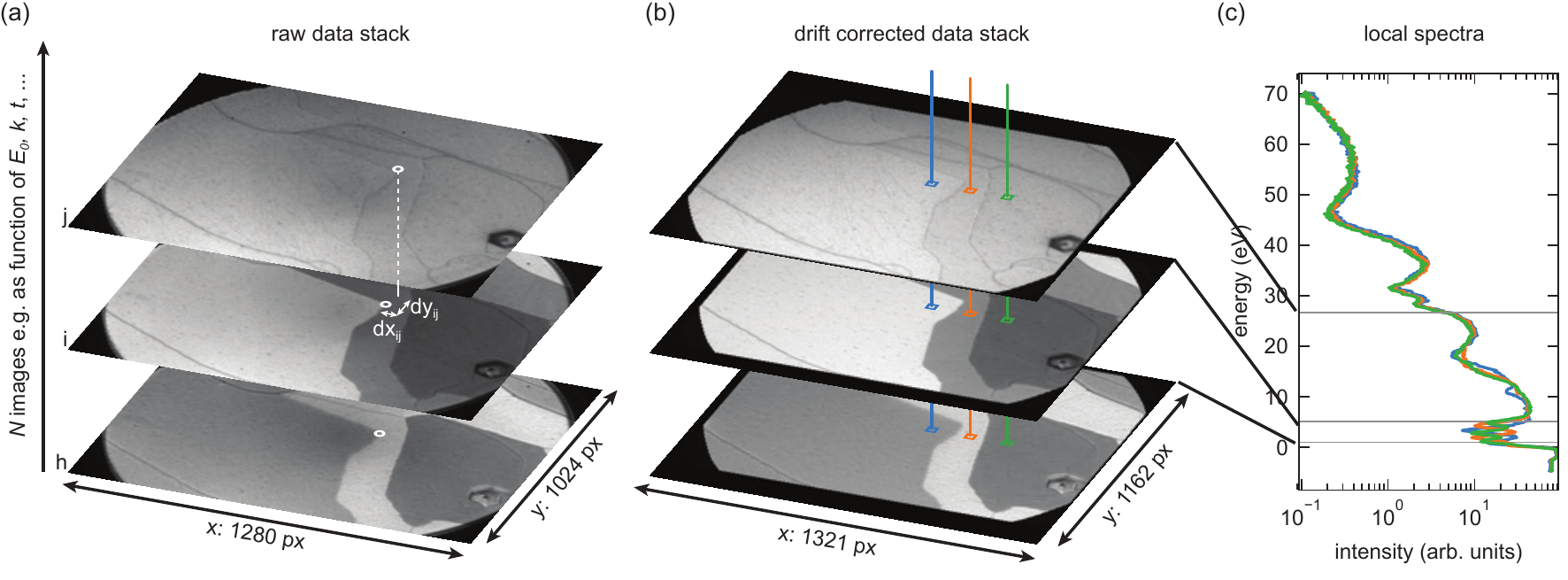}
\caption{\subf{a} A stack of raw LEEM images where images are shifted with respect to each other due to experimental drift. \subf{b} Drift correction aligns features in the images to compensate. \subf{c} Spectra corresponding to pixels indicated in (b).}
\label{fig:datacube}
\end{figure}

However, to unlock the true potential of quantitative analysis of multi-dimensional LEEM data, post-processing of images and combination with meta-data is needed. 
In particular, it is necessary to correct for detector artifacts and image drift and to convert image intensity to physical quantities. 

To this end, we here present a modular data acquisition and analysis pipeline for multi-dimensional LEEM data, combining techniques well established in other fields such as general astronomy or transmission electron microscopy (TEM), that yields high resolution spectroscopic datasets and visualizations thereof.
In particular, we start with the correction of the raw data for detector artifacts using flat field and dark current correction, as illustrated in Fig.\ \ref{fig:datacube}(a). 
Combining these corrections on the images with active feedback on detector gain enables High Dynamic Range (HDR) spectroscopy, which makes it possible to measure spectra over four orders of magnitude of intensity. 
Subsequently, we demonstrate that compensation of detector artifacts also enables drift correction with sub-pixel accurate image registration, yielding a fully corrected data stack (Fig.\ \ref{fig:datacube} (b)). This creates a true pixel-by-pixel spectroscopic dataset, as shown in Fig.\ \ref{fig:datacube}(c), i.e. every pixel contains a reflectivity spectrum of the corresponding position on the sample. 
Finally, we explore the potential for more advanced computational data analysis. We show that by using relatively simple dimension reduction techniques and clustering, these datasets can be intuitively analyzed and visualized, enabling semi-automatic identification of areas with different spectra. 

To demonstrate these features and quantify the accuracy, we apply the drift correction algorithm to artificial data and then apply the full pipeline to a real dataset acquired on the SPECS P90 based ESCHER system in Leiden~\cite{tromp2010new,schramm2011low,tromp2013new}.
The sample of the dataset is few-layer graphene grown by thermal decomposition of Silicon Carbide (SiC)~\cite{emtsev2009towards}, followed by hydrogen intercalation to decouple the graphene from the SiC substrate~\cite{speck-QFMLG,riedl-QFMLG}. 
Bright Field LEEM spectra can be used to distinguish the resulting mixture of bilayer, trilayer and thicker graphene, as interlayer states cause distinct minima in the reflectivity spectra~\cite{hibino2008microscopic,feenstra2013low}.
In addition, the growth process causes strain-induced stacking domains, which can be distinguished using Dark Field LEEM spectra~\cite{hibino2009stacking,dejong2018intrinsic}.
The sample dataset consists of bright-field and dark-field LEEM images of the same area for a range of landing energies (sometimes referred to as LEEM-I(V)).
The dark-field dataset uses a first order diffraction spot and tilted illumination such that the incident beam has the opposite angle to the normal as the diffracted beam, as described in more detail in Ref.\ ~\cite{dejong2018intrinsic}.
The data is available as open data~\cite{dejong2019data} and is interpreted and investigated in detail in Ref.~\cite{dejong2018intrinsic, dejong2018intrinsicdata}.

\section{Detector Correction}
No physical detector system is perfect, i.e. each detector system introduces systematic errors and noise. 
Knowledge of the sources of these imperfections enables the correction of most of them.
The ESCHER LEEM has the classical detector layout: A chevron microchannel plate array (MCP, manufactured by Hamamatsu) for electron multiplication, a phosphor screen to convert electrons to photons and a CCD camera (a PCO sensicam SVGA) to record images of the phosphor screen. 

The CCD introduces artifacts in the form of added dark counts and a non-uniform gain~\cite{VanGastel2009Medipix, widenhorn2002temperature, widenhorn2010exposure, widenhorn2001meyer}.
Furthermore, the MCP gain is also spatially non-uniform, for example due to overexposure damaging of the MCP, resulting in locally reduced gain.
Therefore we describe the measured intensity $I_\text{CCD}$ on the CCD as the following combination of the previously named detector artifacts and the `true' signal $I_\text{in}$:
\begin{equation}
I_\text{CCD}(x,y) = DC(x,y) + I_\text{in}(x,y)\cdot G(x,y)
\label{eq:intensities}
\end{equation}
Where $DC(x,y)$ is the intensity caused by dark current and $G(x,y)$ is the position-dependent and as-of-yet unknown gain factor comprising all modifications to the gain.
\begin{figure*}[!ht]
\centering
\includegraphics{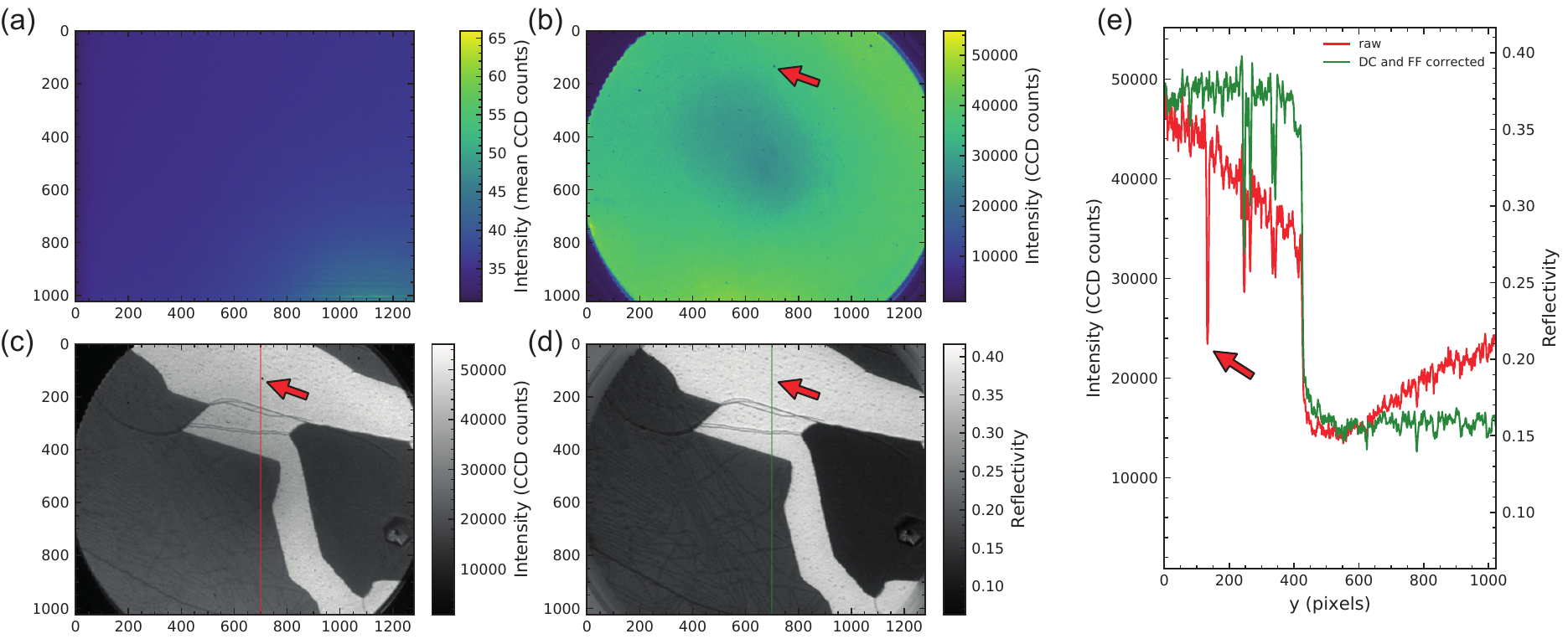}
\caption{\subf{a} Dark Count image taken on ESCHER averaged over $19\times 16$ images with \SI{250}{\milli\second} exposure time. \subf{b} Flat field image with visible edges of the round microchannel plate and damaged areas (arrow). \subf{c} Uncorrected bright-field LEEM image from the sample dataset. The field of view corresponds to \SI{3.5}{\micro\metre}. \subf{d} Dark count and flat field corrected version of the image in (c).
\subf{e} Line cut through the raw image (c) and the corrected image (d) shown in red and green, respectively. Note that the dip due to MCP damage at $y=140$ (arrow) is removed and the profiles for similar areas are flattened.}
\label{fig:detectorcorr}
\end{figure*}

To compensate for these detector artifacts, we employ techniques well-established in astronomy (and other fields using CCD cameras) to effectively invert the relation in Eq.\ (\ref{eq:intensities}), to extract $I_\text{in}(x,y)$ without the deleterious effects of background dark counts $DC(x,y)$ and local gain variations $G(x,y)$. 

First, the dark current of the CCD is compensated by pixel-wise subtracting a non-illuminated \textit{dark count} image, i.e. an image with the same exposure time as used for the measurement, but no electron illumination at all. A pixel-wise average of a set of such dark count images is shown in Fig.\ \ref{fig:detectorcorr}a.
The dark current arises from thermal excitations in the sensor and varies over time with an approximately Gaussian distribution. 
The mean of this distribution is dependent on the pixel, i.e. the $x,y$ location, for example visible in Fig \ref{fig:detectorcorr}(a) as a slight increase in the lower right corner.
To suppress the thermal fluctuations in the template dark current image, it is desirable to average over several dark count images to prevent the introduction of systematic errors.
We assume that the per-pixel dark currents are identically distributed with a variance $\text{Var}_\text{therm}$ except for a spatial variance $\text{Var}_\text{spatial}$ of the mean.
This is mathematically equivalent to assuming the dark current fluctuates around its mean with both spatially dependent (but fixed in time) noise and time-dependent thermal noise.
By averaging multiple dark images, we reduce the thermal variance but not the spatial variance. The remaining variance is given by:

\begin{align*}
\text{Var}_\text{tot}(n) &= \text{Var}_\text{therm}(n )+ \text{Var}_\text{spatial}\\
&= \frac{1}{n}\text{Var}_\text{therm}(1)+ \text{Var}_\text{spatial}
\end{align*}
Where $\text{Var}(n)$ is used to denote the variance of $n$ pixel-wise averaged images.
By determining $\text{Var}_\text{tot}(n)$ and $\text{Var}_\text{tot}(1)$ experimentally we can isolate the thermal noise on a single image:
\begin{equation}
\text{Var}_\text{therm}(1)= \big[\text{Var}_\text{tot}(1 ) - \text{Var}_\text{tot}(n )\big]\cdot \frac{n}{n-1}
\end{equation}

For the ESCHER system with its Peltier-cooled camera, we find $\text{Var}_\text{therm}(16\times \SI{250}{\milli\second}\text{ images}) = 114.3$.
Therefore, a set of $120\times16$ images (a total exposure time of 8 minutes) is sufficient to suppress the systematic errors to values smaller than the discretization error.
We find that the dark count image does not significantly change over time, and therefore remeasuring dark count images is seldomly needed.\\ 

Second, to compensate for spatial gain variations, which are mostly due to the MCP, a (conventional) \textit{flat field} correction is performed, dividing the full dark count-corrected dataset by an evenly illuminated image~\cite{seibert1998flat}.
In LEEM, at negative landing energies of $E_0\approx \SI{-20}{\electronvolt}$ the sample behaves as a mirror, yielding an almost perfect flat field image as approximation of $G(x,y)$ in Eq.\ (\ref{eq:intensities}). 
A relatively large value for the negative energy is taken to prevent artifacts from local in-plane electric field components, e.g. due to work function or height differences in the sample~\cite{yu2010phase, Kennedy-MM-distortions, jobst2018quantifying}.
For the ESCHER system, it is necessary to take flat field images within hours of the measurement, as the MCP wears over time and the gun emission profile and system alignment change on relatively short timescales~\cite{Schramm2012Intrinsic}.
Furthermore, taking a flat field image at the same precise alignment as the measurement is preferred for two reasons. 
First, barring absolutely perfect alignment of the system as well as a perfectly uniform emission from the electron gun, the beam intensity is not spatially uniform. As illumination inhomogeneities are dependent on the precise settings of the lenses, these will also be compensated for if the flat field is recorded in the exact same configuration. 
Second, for proper normalization of the data, as explained in Section \ref{sec:HDR}, the same magnification (projector settings) is needed. 

An alternative to this mirror mode flat fielding is to average over a sufficiently large set of images of different positions on the sample and use the resulting average as a flat field image.
In most cases however, mirror mode flat fielding is preferred over such ensemble-average flat fielding since for the latter many images of different locations are required. 
Even when such a set is already available, it is hard to rule out any systematic (statistical) errors.
Lastly ensemble-average cannot provide proper normalization of the data to convert to true reflectivity.


\section{High dynamic range spectroscopy}\label{sec:HDR}
In LEEM and LEED, large variations occur in the amplitude of the signal, both within individual images and from image to image. For example, in LEED spectroscopy, features of interest are often orders of magnitude less bright than primary Bragg peaks. 
This necessitates a detector system with a large dynamic range. 
The CCD-camera of the ESCHER setup has a bit depth of 12 bits and a possibility to accumulate 16 images in hardware, yielding an effective bit depth of 16 bits for singular images.

For most materials, the reflected intensity $I(E_0)$ changes over orders of magnitude as a function of $E_0$. Starting in mirror mode, the reflected intensity tends to decrease roughly exponentially for $E_0 \lesssim \SI{100}{\electronvolt}$. 
To obtain spectra with such a large dynamic range, the dynamic range offered by the bit depth of the CCD alone is not sufficient.

However, the gain $G$ of the MCP, i.e. the ratio of outgoing electrons to incoming electrons, can be tuned by the voltage $V_\text{MCP}$ applied over the MCP. 
This gain scales approximately exponential in $V_\text{MCP}$ (over a reasonable range, see next section), enabling image formation of approximately constant intensity on the CCD, for a wide range of incident electron intensities. 
We use this property to develop a scheme to further increase the dynamic range in which $G(V_\text{MCP})$ is adjusted by setting a new MCP bias for each new image, i.e. increasing the gain for images where the reflected intensity is low.
Measuring $V_\text{MCP}$ for each recorded image and calibrating $G(V_\text{MCP})$ makes it possible to employ the full dynamic range of the CCD-camera for all landing energies, without losing the information of the absolute magnitude of the measured intensities, thus extending the range of spectroscopy without significant decrease in signal-to-noise ratio.

\subsection{Calibration}
\begin{figure}
\includegraphics[width=\columnwidth]{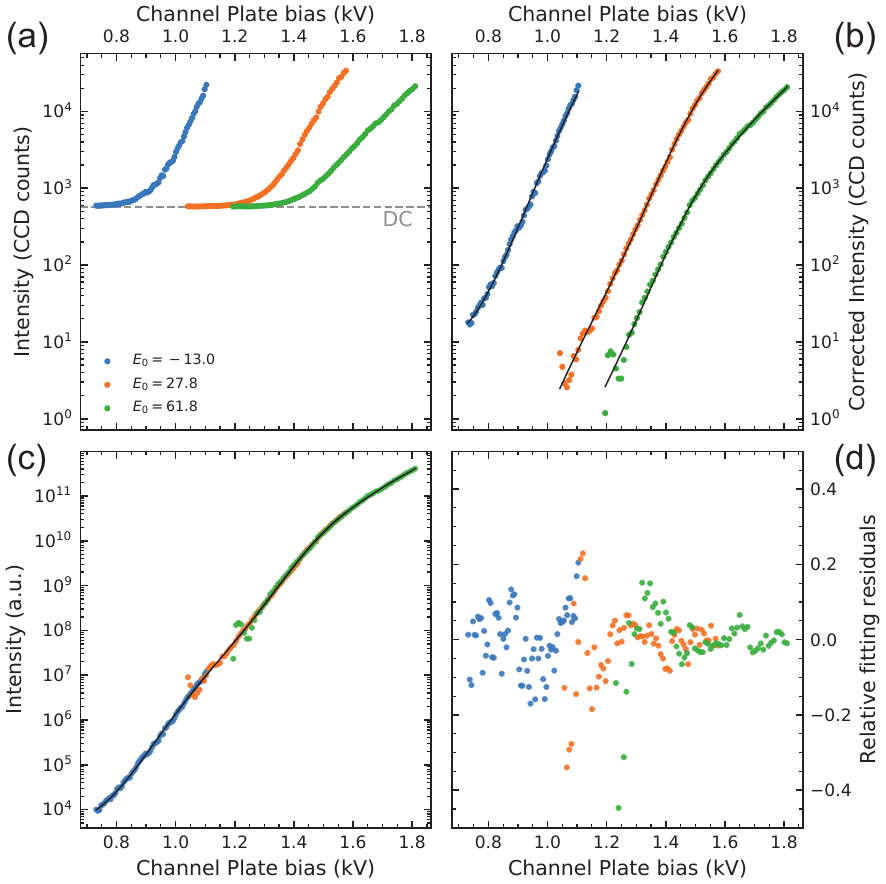}
\caption{\subf{a} Calibration curves as measured with $16\times\SI{250}{\milli\second}$ exposure time per image measured on graphene on SiC. \subf{b} Calibration curves corrected for dark count. \subf{c} Calibration curves with matched intensity and normalized by joint curve fit of Eq.\ (\ref{eq:MCPgain}) and resulting best fit (black line). \subf{d} Residuals of the joint fit in (c).}
\label{fig:calibration}
\end{figure}

Hamamatsu Photonics K.K., the manufacturer of the microchannel plate in the ESCHER setup, specifies an exponential gain as function of voltage for a part of the range of possible biases~\cite{hamamatsuMCP}. 
To extend the useful range beyond this limit and thus enable the use of the full bias range up to the maximal $\SI{1800}{\volt}$
, the gain versus bias curve was calibrated as follows:
\begin{enumerate}
\item First, in mirror mode, $V_\text{MCP}$ is adjusted such that the maximum intensity in the image corresponds to the full intensity on the CCD, staying just below intensities damaging the MCP.
\item While decreasing $V_\text{MCP}$, images are acquired for evenly spaced bias values. The intensities of these images form the dataset for calibration of the low bias part.
\item Returning $V_\text{MCP}$ to the previous maximum value, $E_0$ is increased until the intensity of the image is so low that it is barely distinguishable from the dark count. 
\item Again $V_\text{MCP}$ is turned up until the maximum image intensity corresponds to the maximum CCD intensity.
\item Steps 2.\ to 4.\ are repeated until a dataset is acquired starting at maximum MCP bias $V_\text{MCP}$.
The resulting curves are shown in Fig.\ \ref{fig:calibration}(a).
\item These datasets are then corrected for dark count as discussed above, resulting in the curves shown in Fig.\ \ref{fig:calibration}(b). Comparing to the uncorrected curves, the increase in accuracy for low intensity values, crucial for accurate calibration, is very apparent.
\item A joint fit of Eq.\ (\ref{eq:MCPgain}), allowing for a different amplitude $A_i$ for each curve, is performed to the corrected data to obtain a general expression for MCP gain $G$ as a function of $V_\text{MCP}$. 
The fit is performed using least squares on the logarithm of the original data with no additional weights, to ensure a good fit over the large range of orders of magnitude.
The fitted curve is then normalized to a convenient value, e.g. $G(1\,\text{kV}) = 1$. 
This normalization can be freely chosen, as $G$ will be applied equally to datasets and flat field images, yielding absolute reflectivity as resulting data.
\end{enumerate}

A first choice for a fitting function would be a simple exponential, but this would not account for any deviation from perfect exponential gain, visible as deviations from a straight line in Fig.\ \ref{fig:calibration}(b). 
For the ESCHER setup we therefore choose to add correction terms of odd power in the exponent:
\begin{equation}
    G(V_\text{MCP}) = A_i\exp\left(\sum_{k=0}^5 c_k {V_\text{MCP}}^{2k+1}\right)
    \label{eq:MCPgain}
\end{equation}
Only odd powers were used to accurately capture the visible trends in the data.
For the ESCHER setup correction terms up to order $V_\text{MCP}^{19}$ ($k< 9$) turned out to give a satisfactory good approximation, as illustrated by the residuals in Fig.\ \ref{fig:calibration}(d).

\subsection{Active per-image optimization of MCP bias}
The resulting curve with calibration coefficients is then used to actively tune the MCP bias during spectroscopic measurements:
A desired range is defined for the maximum intensity on the camera, corresponding to a maximum safe electron intensity on the MCP to prevent damage on the one hand, and a minimum desired intensity of the image on the CCD on the other hand.
Whenever the maximum intensity of an image falls outside this range, the MCP gain $G(V)$ will be adjusted such that the intensity of the next image again falls in the center of this range.
Assuming the intensity changes continuously, this method ensures the use of the full intensity range of the camera for each image, while protecting the MCP against damage.

\begin{figure}[!h]
\includegraphics[width=\columnwidth]{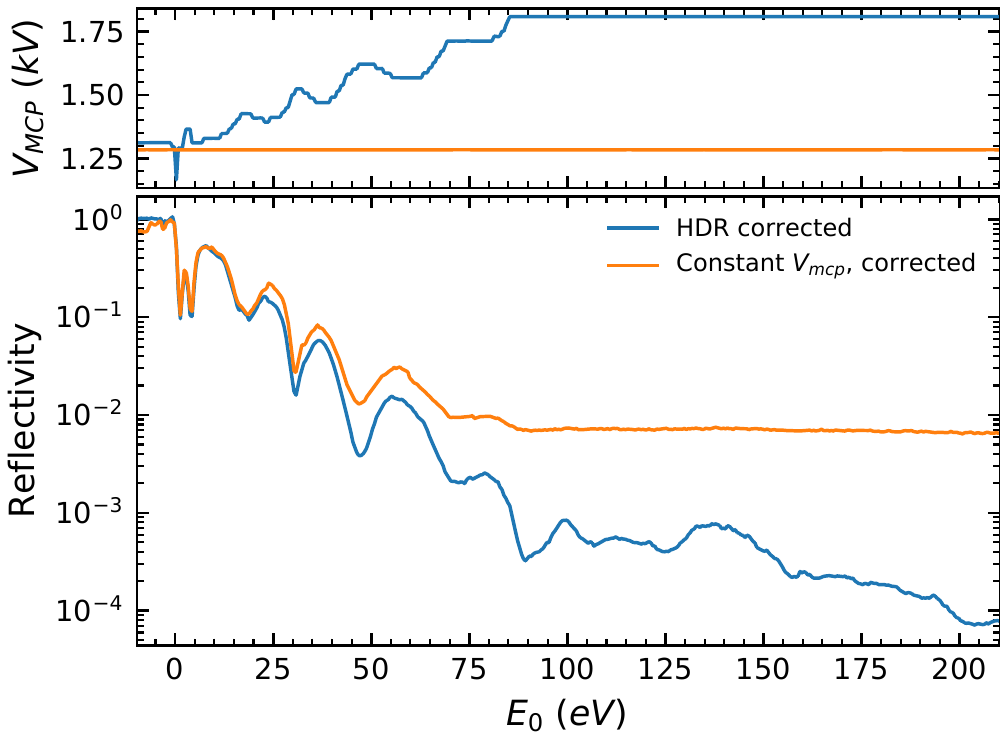}
\caption{Regular spectroscopic reflectivity curve (orange) of bilayer graphene on SiC, corrected for dark count and flat field, but with a single setting of $V_\text{MCP}$ (top panel). The HDR measurement of the same area with active MCP bias tuning (blue) can resolve details down to lower intensity.}
\label{fig:HDRcomparison}
\end{figure}
Additionally, after the measurements, the calibration curve is used to calculate the real, relative intensity from the image intensity and the recorded $V_\text{MCP}$.
By dividing this intensity by the intensity of the flat field image (taken in mirror mode and corrected for dark current and the MCP bias), we calculate a (floating point) conversion factor to true \textit{reflectivity} values for each image. 
These ratios are added to the metadata of every image.
By applying this conversion as a final step after any analysis of the data, errors due to discretization of highly amplified, and therefore low true intensity, images are minimized. 
Note that this procedure makes the conversion to true reflectivity possible even for datasets with no mirror mode in the dataset itself, such as dark field measurements.

\subsection{Comparison of results}

Spectroscopic LEEM-I(V) curves on bilayer graphene on silicon carbide are measured both with constant MCP bias and with adaptive MCP bias as described above. 
A comparison between the resulting curves is shown in Fig.\ \ref{fig:HDRcomparison}.
While the regular, constant MCP, curve starts to lose detail around $E_0=\SI{50}{\electronvolt}$, i.e. after a factor of 100 decrease in signal, the adaptive measurement captures intensity variations in the spectrum almost 4 orders of magnitude lower than the initial intensity.
We thus call the adaptive method high dynamic range (HDR) imaging. 


\section{Drift Correction by image registration}

In LEEM imaging, the position of the image on the detector tends to shift during measurement as shown in Fig.\ \ref{fig:datacube}(a). This prevents per-location interpretation of the data, both for spectroscopic measurements and measurements with varying temperature. 
Although the shift can be minimized by precise alignment of the system, we find that a significant shift always remains, especially in tilted illumination experiments such as DF-LEEM or angle-resolved reflected-electron spectroscopy (ARRES)~\cite{jobst2015nanoscale, jobst2018quantifying, dejong2018intrinsic}, which makes the compensation of this image drift necessary.

This problem has been studied in depth in the field of image registration, motivated by wide-ranging applications such as stabilization of conventional video, combination of satellite imagery and medical imaging~\cite{Foroosh2002,Klein2010elastix,shamonin2014fast}. 
Techniques generally rely on defining a measure of similarity between a template and other images, either by some form of cross correlation, or by identifying specific matching features in both. 
The image is then deformed by a fixed set of transformations (either affine, i.e. purely shifts and rotations or non-rigid, i.e. additional deformation), until the match between the features in the images and the features in the template is maximal.
For LEEM data, the measurement drift is almost completely described by in-plane shifts, significantly reducing the space of expected transformations. 
A common approach in this case is to use the (two-dimensional) cross-correlation as a measure of similarity between two shifted images and to find the maximum for all images compared to a template, as the location of maximum of the cross-correlation corresponds directly to the shift between the image and the used template. 

The cross correlation of two $n\times n$ pixels images $I_1(x,y)$ and $I_2(x,y)$ is defined as follows:
\begin{equation}
    \mathcal{C}(I_1,I_2)(x,y) = \frac{1}{n^2}\sum_{x'=0}^{n-1}\sum_{y'=0}^{n-1} I_1(x',y')I_2(x+x',y+y')
\label{eq:Corrdef}
\end{equation}
where the coordinates can be wrapped around, i.e. all spatial coordinates are modulo $n$. Furthermore, we can relate this to the convolution operation (denoted as $\circ$):
\begin{equation}
\begin{aligned}
    \mathcal{C}(I_1,I_2)(x,y) &= \frac{1}{n^2}\sum_{x'=0}^{n-1}\sum_{y'=0}^{n-1} I_1(x',y')I_2(x - (-x'), y - (-y'))\\ 
    &=: \big(I_1(x',y') \circ I_2(-x',-y')\big)(x,y)
\end{aligned}
\end{equation}
Using this, the cross correlation can be expressed in terms of (two-dimensional) Fourier transforms $\mathcal{F}$:
\begin{equation}
    \mathcal{C}(I_1,I_2) = I_1(x',y') \circ I_2(-x',-y') = \mathcal{F}^{-1}\left(\mathcal{F}(I_1)\cdot \overline{\mathcal{F}(I_2)}\right)
    \label{eq:corrandffts}
\end{equation}
Where $\overline{\mathcal{F}(I_2)}$ denotes the complex conjugate of the Fourier transform of image 2.
This makes the cross-correlation extra suitable as a measure of similarity, since it can be computed efficiently using the two-dimensional Fast Fourier Transform (FFT).
Determining the local maximum of the cross-correlation yields the integer shift for which the two input images are most similar, with the height of the maximum an indication of the quality of the match.
To further increase accuracy, several variants, such as gradient cross-correlation and phase-shift cross-correlation, have been shown to achieve sub-pixel accuracy for pairs of images~\cite{Foroosh2002,IsmailiAalaoui2008,guizar2008efficient,Tzimiropoulos2011}.

\begin{figure}[!ht]
\centering
\includegraphics[width=\columnwidth]{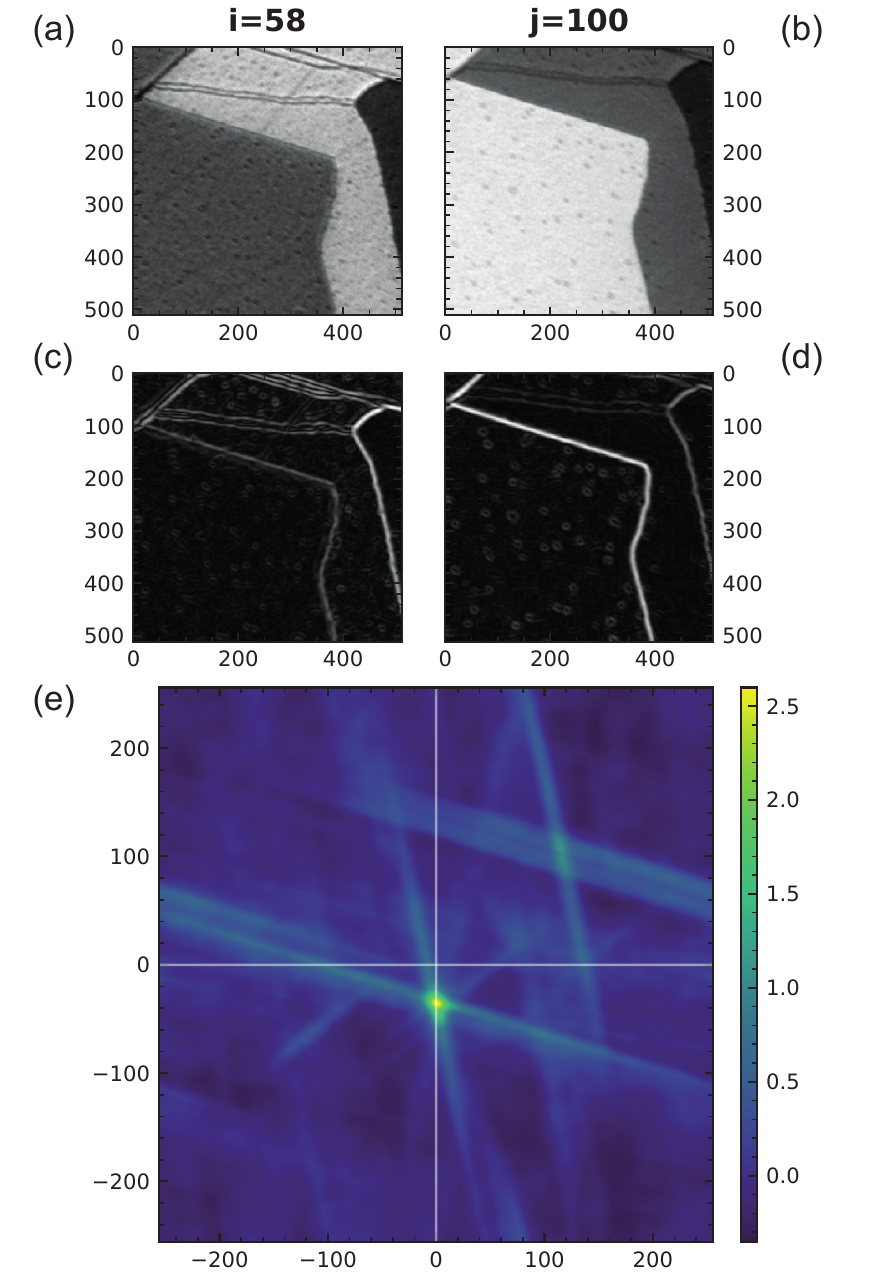}
\caption{\subf{a,b} Two bright-field LEEM images of few-layer graphene obtained at different $E_0$. \subf{c,d} Their Gaussian and Sobel-filtered versions with a Gaussian standard deviation of $3$ pixels highlights the edges and erases the contrast inversion. \subf{e} The cross-correlation of the filtered images exhibits a clear maximum. Its position compared to zero (white lines) corresponds to the relative shift of the images.}
\label{fig:filtering}
\end{figure}

For LEEM data however, the straightforward cross-correlation approach is often hindered by the physics underlying the electron spectra, resulting in contrast changes (c.f.\ Fig.\ \ref{fig:filtering}(a,b)) and even inversions for different values of $E_0$.
The problem can be slightly alleviated by using multiple templates, but in general this approach is unsatisfactory. 
Instead we present another approach here: We first apply digital filters and then compare each image to all other images, similar to the algorithm by Schaffer et al.\ for energy filtered transmission electron microscopy~\cite{Schaffer2004Automated}. 
It then uses a statistical average of the found integer shifts between all pairs of images to achieve sub-pixel accuracy.

We analyze the accuracy of this algorithm using an artificial test dataset and show that the accompanying Python implementation~\cite{tobias2019github} is fast enough to process stacks of hundreds of images in mere minutes by performing benchmarks on a real dataset, followed by a discussion of the algorithm and results.\\

\begin{figure}[!ht]
\centering
\includegraphics[width=\columnwidth]{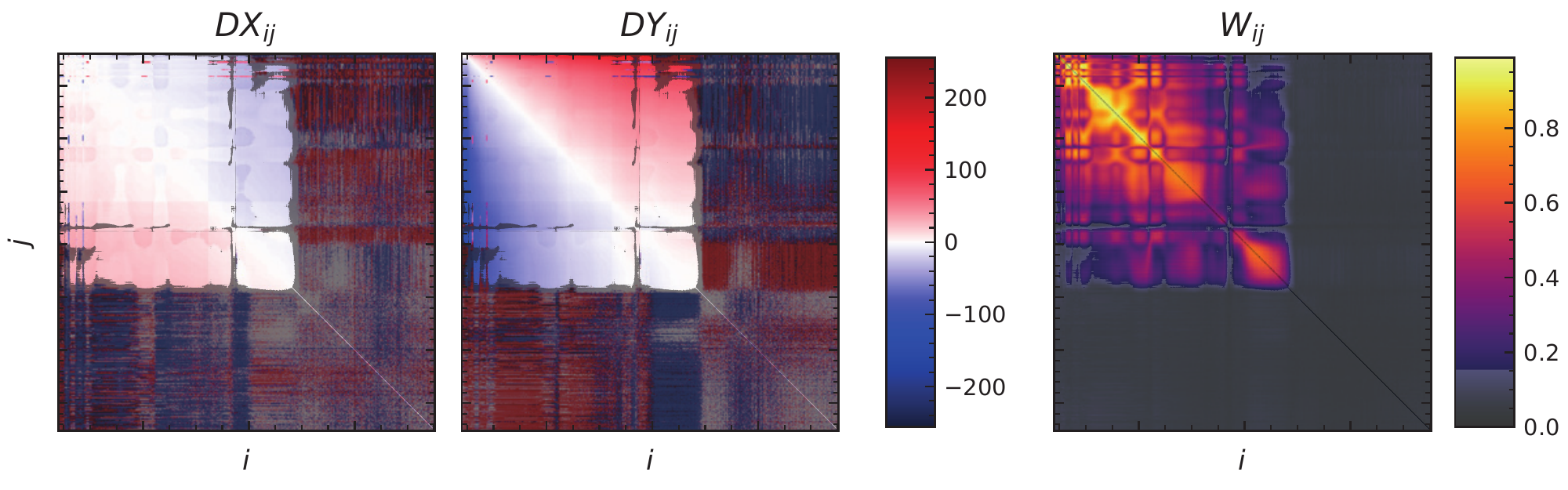}
\caption{Calculated shift matrices $DX$ and $DY$ and weight matrix $\overline{W}_{ij}$ for the bright-field dataset. Matches of a weight below $W_{\min} = 0.15$ (shaded in gray) are mostly false positives. Consequently, they are set to zero weight in the algorithm.}
\label{fig:weights}
\end{figure}

The algorithm consists of the following steps:
\begin{enumerate}
\item Select an area of each of the (detector-corrected) $N$ images, suitably sized for FFTs (i.e. preferably $2^n \times 2^n$ pixels).
\item Apply Gaussian smoothing with standard deviation of $\sigma$ pixels to reduce Poissonian noise in the images.
\item Apply a (magnitude) Sobel filter to highlight edges only, as shown in Fig.\ \ref{fig:filtering}(c) and (d). As such, images with inverted contrast (c.f. Fig.\ \ref{fig:filtering}(a) and (b)) become similar to each other.
\item Using Eq.\ (\ref{eq:corrandffts}), compute the cross-correlation, as shown in Fig.\ \ref{fig:filtering}(e). Do this for all pairs $(i,j)$ of images.
\item Compute the location $(DX,DY)_{ij}$ and value $W_{ij}$ of the maximum of the cross-correlation for all image pairs $(i,j)$. $DX_{ij}$ and $DY_{ij}$ form the anti-symmetric matrices of found relative shifts in either direction, while $W_{ij}$ is a symmetric matrix of weights of the found matches, as shown in Fig.\ \ref{fig:weights}.
\item Normalize the maximum values $W_{ij}$ to be used as weights in step 8: $\overline{W}_{ij} = \frac{W_{ij}}{\sqrt{W_{ii}\cdot W_{jj}}}$.
\item Pick a threshold $W_{\min}$ to remove any false positive matches between images. A threshold of $W_{\min}=0.15$, based on $DX$, $DY$ and $W_{ij}$, is shown in Fig.\ \ref{fig:weights} as gray shading. Set $\overline{W}_{ij}=0$ for all $\overline{W}_{ij}<W_{\min}$.
\item To reduce the $N^2$ relative shifts $DX$ to a length $N$ vector of horizontal shifts $dx$, minimize the errors $\left(dx_i - dx_j - DX_{ij}\right) \overline{W}_{ij}^{4}$
(using least squares). Do the same with $DY$ to obtain the vertical shift vector $dy$.
\item Apply these found shifts $dx$ and $dy$ to the original detector corrected images, interpolating (either bi-linearly or via Fourier) for non-integer shifts.
\end{enumerate}

\subsection{Accuracy testing}
\begin{figure}
\centering
\includegraphics[width=\columnwidth]{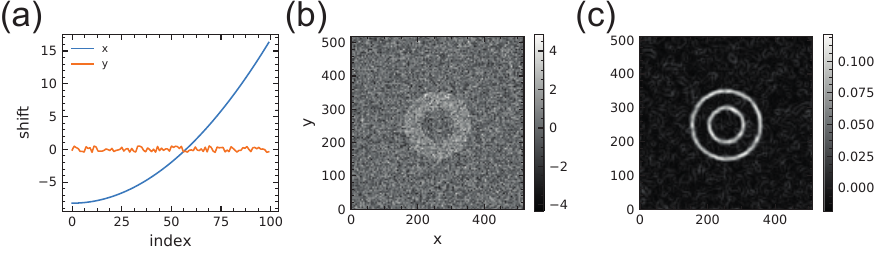}
\caption{\subf{a} Image shifts in $x$ and $y$ used for the synthetic dataset. \subf{b} Image 0 of the synthetic dataset with Gaussian noise with standard deviation of $A=1.0$. \subf{c} Gaussian- and Sobel-filtered version of (b), with a Gaussian filter width of 5 pixels, highlighting edges.}
\label{fig:simdata}
\end{figure}
To validate and benchmark the accuracy of the drift correction algorithm beyond visual inspection of resulting drift corrected datasets, an artificial test dataset with known shifts was created. This enables exact comparison of results to a `true' drift.

The test dataset, as shown in Fig.\ \ref{fig:simdata}, consists of $N=100$ copies of an annulus of intensity $1.0$ on a background of $0.0$. The dataset is shifted over a parabolic shift in the $x$-direction and random shifts uniformly chosen from the interval $[-0.5,0.5]$ pixels in the $y$ direction (see Fig.\ \ref{fig:simdata}(a)). Finally pixel-wise Gaussian (pseudo-) random noise is added to all images. The standard deviation $A$ of the added random noise is then varied to simulate images with different signal-to-noise ratios (SNR).

\begin{figure}
\centering
\includegraphics[width=\columnwidth]{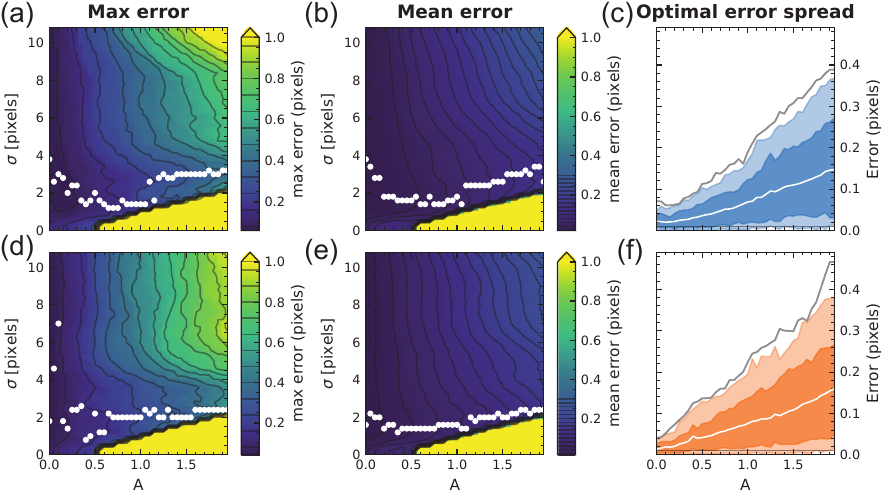}
\caption{\subf{a,b} Maximum and mean error in $dx$ shift as calculated by the algorithm for different values of noise amplitude $A$ and smoothing parameter $\sigma$. The optimal value of the Gaussian smoothing $\sigma$ as a function of added noise amplitude $A$ is drawn in white. Black contour lines are added as a guide to the eye.
\subf{c} Spread of the error for the optimal values of $\sigma$ for varying $A$. Dark and light bands are respectively $1$ and $2$ standard deviations, maximum error is indicated as gray line.
\subf{d,e,f} Same for the $y$ direction.}
\label{fig:simulationresults}
\end{figure}

The resulting maximum error in the found shift compared to the original, `true' shift, as well as the resulting mean error for different values of $A$ and $\sigma$ is shown in Fig.\ \ref{fig:simulationresults}, separately for the $x$ and $y$ directions.
These results verify that, at least for synthetic datasets, the algorithm achieves sub-pixel accuracy, with the mean absolute error in pixels of about 0.1 times the relative noise amplitude $A$ for the optimal value of smoothing $\sigma$, and the maximum absolute error just reaching 0.5 pixel for the extreme value of $A=2$. 
As expected, the error is strictly increasing for decreasing SNR, i.e. increasing $A$. 
After an initial cutoff, visible in saturated yellow in Fig.\ \ref{fig:simulationresults}, the accuracy of the algorithm is also generally decreasing for increasing smoothing width $\sigma$. However, after this initial cutoff, there is a comfortably large range of $\sigma$ where the algorithm performs well.

We found that most features visible in the high-$\sigma$, high-$A$ regime of Fig.\ \ref{fig:simulationresults} (b,d) are dependent on the initialization of the random generator for the added pixel-wise noise and are thus not significant (c.f.\ a second run in Supplementary Fig.\ \ref{suppfig:simulationresults}).

\subsection{Time complexity}
\begin{figure}[!ht]
\centering
\includegraphics[width=\columnwidth]{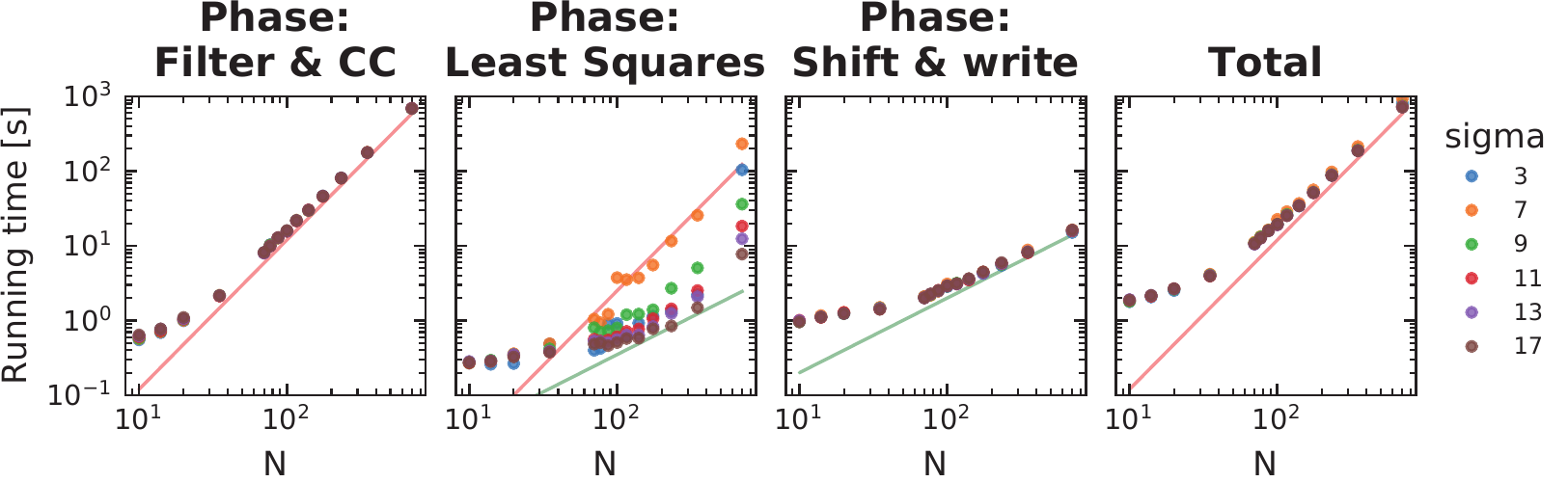}
\caption{Run times of different phases of the drift correction algorithm on $256\times256$ pixel images. Linear (green) and quadratic (red) slopes are added as a guide to the eye, illustrating that the cross correlations scales quadratically in the number of images $N$, while the shifting and saving of images itself is linear.}
\label{fig:timebench}
\end{figure}

To benchmark the computational complexity of the algorithm, it was applied to subsequently larger parts of the real dataset, while measuring the computation times for the least squares optimization (step 8 above) and the shifting and writing of images (step 9) separately. 
The results show calculating the cross-correlations takes the most time, as it scales almost perfectly quadratically in the number of images $N$, as shown in Fig.\ \ref{fig:timebench}. 
The shifting and writing of images scales linearly and is not significant for larger datasets. The total time therefore scales nearly perfectly quadratically, with a dataset of 500 images drift corrected in less than 7 minutes of computation time (exact details of the hardware and software used for benchmarking can be found in the Supplemental Information).
As such, LEEM spectroscopy datasets can be comfortably and regularly drift corrected on a desktop PC.

\subsection{Discussion}
We elaborate here on the choices made in the algorithm.
The use of the magnitude Sobel filter has multiple benefits, similar to using the gradient cross-correlation: Contrast inversions between areas with different spectroscopic properties nonetheless result in similar images (c.f.\ Fig.\ \ref{fig:filtering}(c,d)). In addition, the constant zero background reduces errors due to wrap-around effects due to performing the calculation in the Fourier domain.

The exponent 4 for the weighing matrix $W_{ij}$ in the least squares minimization step 8 was empirically found to give the best results for real datasets.

As already noted by Schaffer et al., the use of cross-correlation between multiple image pairs and combining the returned integer shift values enables sub-pixel accuracy. The maximum theoretical accuracy is $\frac{1}{N} \text{ pixels}$, but is reduced for pixels where false positive matches are thresholded out.

Alternative methods of obtaining sub-pixel accuracy in determining shifts include a combination of upscaling and matrix-multiplication discrete cosine transforms~\cite{guizar2008efficient} and a rather elegant interpolation of the phase cross-correlation method proposed by Foroosh et al.~\cite{Foroosh2002}.
However our current method is less complex and combines robustness against global drift with handling of changing contrasts, which is crucial for spectroscopic LEEM data. 
Although the sub-pixel precise phase correlation method seems a straightforward extension of regular cross correlation, it is less suitable for datasets with changing noise levels and not suitable for false positive detection by normalization, both properties we found essential to handle spectroscopic LEEM data.

Contrary to Schaffer et al., we found smaller values of Gaussian smoothing width $\sigma$ yield the best results, with larger values yielding artificial shifts around contrast inversions for real datasets and generally performing worse for the synthetic dataset, as visible in Fig.\ \ref{fig:simulationresults}. 

Schaffer et al. found their approach at the time (2004) not computationally feasible for large amounts of images but, as shown in the previous section, the current implementation is able to drift correct a stack of several hundreds of images comfortably on a single desktop computer.
We want to emphasize that the use of \texttt{Python} gives flexibility and makes it easy to adapt the code. For instance, increasing performance even more lies within reach by performing the FFT cross-correlations and maximum search on one or more graphical processing units using one of several libraries or by using a cluster running a \texttt{dask} scheduler.
Further speedup would be possible by pruning which pairs of images are to be cross-correlated.
An avenue not explored here, is the use of pyramid methods to create a multi-step routine where firstly a fast estimate of the shift is computed on a smoothened and reduced-size image before using consecutively larger images to refine the estimate~\cite{adelson1984pyramid,thevenaz1998pyramid}.

Beyond drift correction, the same method presented here can also be applied to create precisely stitched overview images of areas much larger than the electron beam spot size. Although, as no contrast inversions or large feature differences are expected for the matching areas, the added value of using a gradient filter is nullified. 
Additionally the number of images that can be matched to the same template is limited, forcing a low upper bound on the sub-pixel accuracy of the optimization part of the algorithm.
Instead, we found that an algorithm based on more regular phase-shift cross-correlation is sufficient for sub-pixel accurate stitching.

\section{Dimension reduction}\label{sec:dimred}

\begin{figure*}[!ht]
\centering
\includegraphics{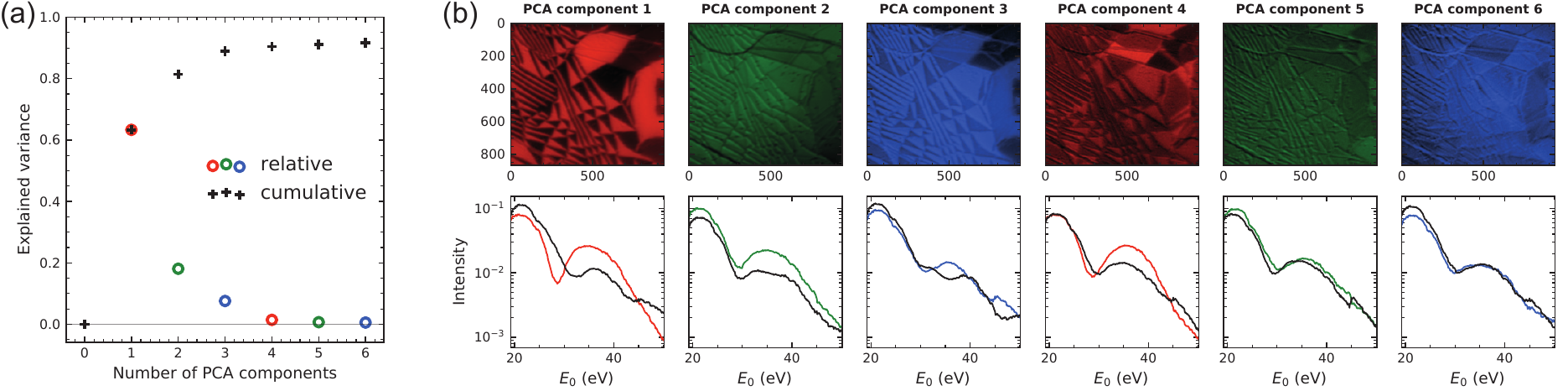}
\caption{\subf{a} Scree plot indicating the retained variance per PCA component for the Dark Field dataset.
\subf{b} Images of the first six PCA components for the Dark Field sample dataset and the spectra corresponding to the maximum and minimum of the respective components occurring in the dataset.}
\label{fig:dimreduc}
\end{figure*}

The sub-pixel accuracy drift correction now makes it possible to reinterpret a LEEM-I(V) dataset as a truly per-pixel set of spectroscopic curves, opening up possibilities for further data analysis.
For a dataset of $N$ images, each such curve (c.f.\ Fig.\ \ref{fig:datacube}) can be seen as a vector in a $N$-dimensional vector space of the mathematically possible spectra. Even for moderate datasets of a few hundred images this is a huge vector space.
In almost all cases however, the physical behavior of the data can be described with a model with far fewer degrees of freedom, i.e.\ the vector space of physically possible spectra has a much lower number of dimensions. 
Therefore, it should be possible to summarize all significant behavior in a much smaller dataset, which can be analyzed (and visualized) much more easily.

Here, we use Principal Component Analysis (PCA), a linear technique based on singular value decomposition (SVD), often used for dimension reduction in data science fields \cite{wold1987principal,Jesse2009PCA-STM,abdi2010principal}. 
The randomized iterative variant of PCA allows for efficient computation of the largest variance components without performing the full SVD decomposition~\cite{halko2011finding}.
This technique therefore projects the spectroscopic data to a lower dimensional subspace, in such a way that maximal data variance is retained. 
It does so in a computationally efficient way, making it well suited for, and popular in, data science.
Before applying PCA, we crop the dataset to remove any areas that lie outside of the detector for any image inside the used range of $E_0$. 
Additionally, each image is scaled to zero mean and unit variance, to not let brighter images contribute stronger to the analysis as they have larger variance.
A lot of other choices for standardization of the data are possible, most of them with useful results, but for the scope of this paper we adhere to this standard choice.

After performing PCA, the lower dimensional subspace or PCA-space, is now spanned by orthogonal `eigen-spectra', referred to as PCA components.
Since the projection map onto this subspace retains most of the variance in the dataset, it is possible to build an approximate reconstruction of the full physical spectra from the reduced PCA spectra.

For spectroscopic LEEM datasets, we find that reducing down to 6 dimensions is often enough to capture more than 90\% of all variance. This is shown in a so-called scree plot in Fig.\ \ref{fig:dimreduc}(a) for the sample dataset of dark-field images of $N=300$ energies.
The dataset can be projected onto a single PCA component by taking the per-pixel inner product with the corresponding `eigen-spectrum`. This yields images visualizing the variance retained by the respective components, as shown in the top half of Fig.\ \ref{fig:dimreduc}(b) for all 6 PCA components. Below each image, the spectra corresponding to the pixels with the minimum and maximum value of this projection are shown in black and color, respectively.

\subsection{Visualization}

\begin{figure}[!ht]
\centering
\includegraphics[width=\columnwidth]{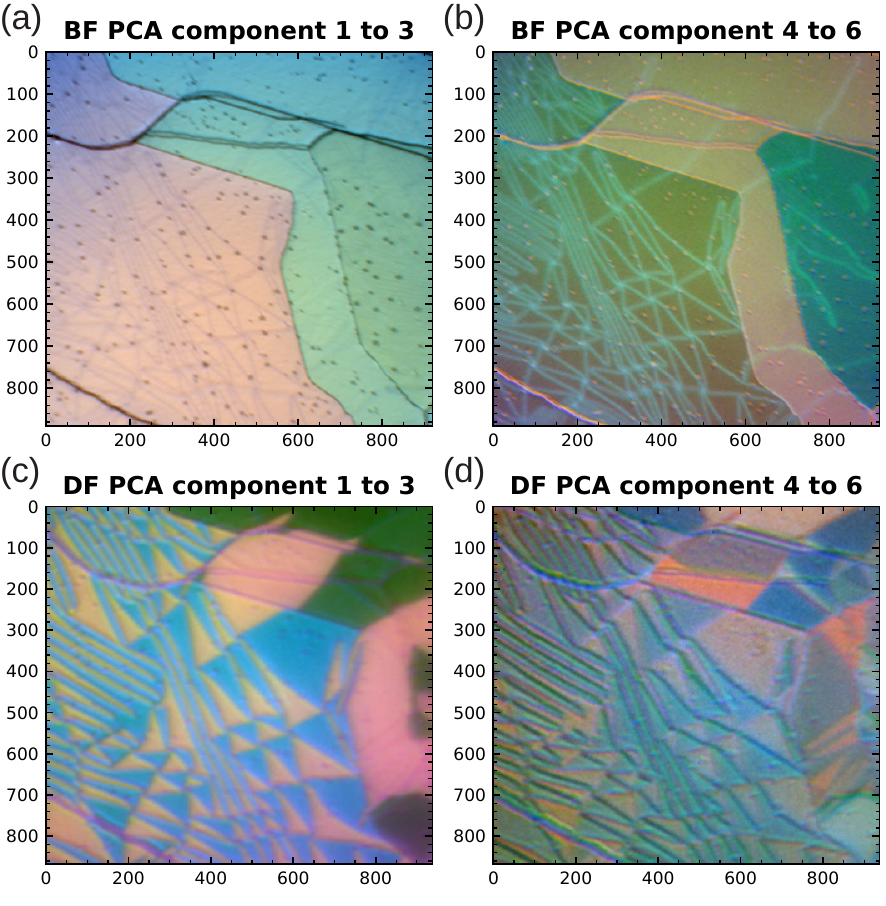}
\caption{The first six PCA components can be used to summarize a spectroscopic LEEM measurement in two RGB pictures. \subf{a,b} Visualization of a spectroscopic bright-field LEEM measurement of quasi-freestanding few-layer graphene on SiC. Different layer counts, stacking boundaries of two types and point defects are distinguishable. \subf{c,d} Visualization of a spectroscopic dark-field LEEM measurement of the same area. All six different possible stacking orders for up to trilayer graphene are easily distinguishable.}
\label{fig:visualization}
\end{figure}

Reducing a spectrum from hundreds of dimensions to a few opens up new opportunities for data visualization.
In particular, it allows for the visualization of nearly all of the variation in spectra of an entire dataset in only two images, as shown in Fig.\ \ref{fig:visualization}.
Here, the values of the six principal components (c.f.\ Fig.\ \ref{fig:dimreduc}), are displayed as the RGB color channels of two pictures per dataset.
To lift the degeneracy in the possibilities of the sign of the PCA components (a PCA eigen-spectrum with the opposite signs retains as much variance), we change the signs such that the a positive projection onto the PCA component corresponds to being brighter in the majority of the images. This way, areas that are bright in the majority of the original images also appear bright in the visualization.
To compensate for the human eyes' preference for green, a scaling of colors as proposed by Kovesi is applied~\cite{kovesi2015good}. It is given by the following matrix:

\[
	\left(
	\begin{array}{c}
		R'\\
		G'\\
		B'\\
	\end{array}
	\right)
	=
	\left(
	\begin{array}{ccc}
		0.90 & 0.17 & 0.00 \\
		0.00 & 0.50 & 0.00\\
		0.10 & 0.33 & 1.00 \\
	\end{array}
	\right)
	\left(
	\begin{array}{c}
		R\\
		G\\
		B\\
	\end{array}
	\right)
\]

\begin{figure*}[!h]
\centering
\includegraphics{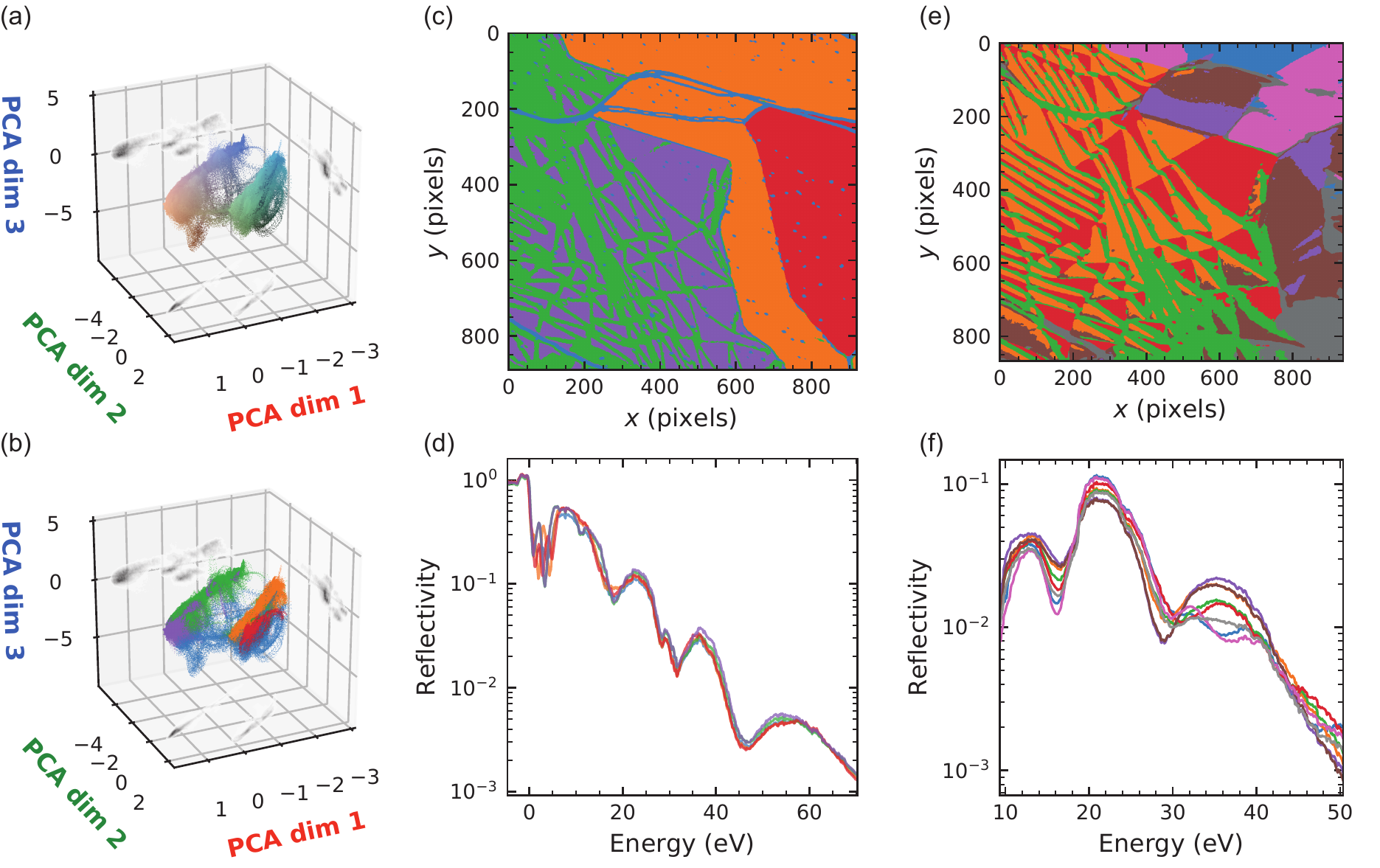}
\caption{\subf{a} Bright-field dataset visualized as point cloud in the space of the first three PCA components. The points are colored according to the mapping in Fig.\ \ref{fig:visualization}(a) and are projected onto the planes in gray scale.
\subf{b} Point cloud as in subfigure (a), but colored according to the computed clustering.
\subf{c} Indication of the cluster labels in the real-space image. \subf{d} Mean bright-field spectroscopy curves for each cluster, automatically recovering layer count and domain walls.
\subf{e,f} Same as subfigures (c,d) for the dark-field dataset reveals all stacking orders present as well as two sets of edge-case curves. C.f.\ Fig.\ 2(c,d) of Ref.~\cite{dejong2018intrinsic}.}
\label{fig:clustering}
\end{figure*}

The results are striking. 
All the sample features are directly visible in Fig.\ \ref{fig:visualization}:
In the bright field dataset, bilayer and thicker graphene are clearly separated in orange and green, respectively in the first three PCA components [Fig.\ \ref{fig:visualization}(a)]. Moreover, SiC step edges, domain walls and point-like defects are clearly visible. 
The next three PCA components [Fig.\ \ref{fig:visualization}(b)] highlights the difference between bilayer (green), trilayer (orange) and four-layer graphene (dark green) and in addition separates step edges (orange), domain boundaries (turquoise) and the defects in different colors. 
Furthermore, two types of domain boundaries can be observed in the four-layer area that are hard to tell apart in conventional LEEM images. The light green and dark ones are presumably domain boundaries in the top-most and lower layers, respectively.

This visualization using the first three PCA components of the dark field dataset [Fig.\ \ref{fig:visualization}(c)] clearly separates the different stacking orders in bilayer (AB in orange and AC in blue) and trilayer graphene. 
The PCA components 4 to 6 [Fig.\ \ref{fig:visualization}(d)] highlights the different stacking orders in trilayer and four-layer areas (different shades of orange and blue) and display an interference effect causing double lines at one type of (corresponding to one direction of) domain edge. 
This clear visualization is particularly remarkable as the dark-field dataset presents a worst case scenario due to its extreme off-axis alignment (see ~\cite{dejong2018intrinsic} for full details), which causes strong image drift and relative shifts of features.

\subsection{Clustering and automatic classification}
In addition to the visualization possibilities explored in the previous section, the dimension reduction by PCA lowers the complexity of the data enough to enable the use of other, more quantitative data analysis techniques. 
In particular, reduction to less than ten dimensions is enough to perform unsupervised classification or clustering on the entire dataset.
Here, we show that a relatively simple clustering algorithm, the classical $k$-means, also known as Lloyd's algorithm~\cite{lloyd1982least}, applied to the PCA reduced dataset, can already be used to distinguish the relevant, different areas.
The structure of the bright field dataset is visualized in terms of the first three PCA components in Fig.\ \ref{fig:clustering}(a), both as a point cloud with colors corresponding to Fig.\ \ref{fig:visualization}(a) and as density projections (gray scale) on the three planes.
The resulting classification from the application of $k$-means to the six PCA components is visualized in the same way in Fig.\ \ref{fig:clustering}(b), where the color of the points now corresponds to the assigned labels.
These same label colors are shown in real space in Fig.\ \ref{fig:clustering}(c).  
In the real space visualization it is clear that the different layer counts are separated (bilayer, trilayer and four-layer as purple, orange and red, respectively) from a class with the point defects and step edges (blue) and a class containing the domain boundaries in the bilayer (green).
The cluster labels can now be used to calculate spectra of each area, e.g. all trilayer pixels without the defects. For this, we take the mean over all pixels belonging to one cluster for each energy. This can be done even for energies outside the range used for the initial clustering as well as for energies where we only have partial data due to drift. The resulting spectra for the clustering in Fig.\ \ref{fig:clustering}(c) are plotted in Fig.\ \ref{fig:clustering}(d).

The same clustering method is applied only to the first 4 PCA components of the dark-field dataset since component 5 and 6 show virtually no distinguishing features and corresponds to very little variance [cf.\ Fig.\ \ref{suppfig:PCAcomponents}(a,b)]. The resulting real space labeling of the clusters and the spectra are shown in Fig.\ \ref{fig:clustering}(e) and (f), respectively.
Here, although not perfect, the clustering algorithm manages to mostly separate the different possible stacking orders (red and orange for the bilayer and purple, brown, pink and blue for the trilayer). The green and gray areas correspond to areas where clear classification as a stacking order is not possible due to phase contrast and non-uniform illumination artifacts due to the tilted illumination. 

Thanks to the proper calibration and mutual registration of the data, this relatively simple algorithm classifies the areas in the dark-field data set with only minor errors (e.g.\ the incorrect assignment of brown trilayer spectrum in the lower left), \emph{without} any input of the positions of each spectrum in the image or any input about the expected differences between spectra.
We anticipate that this classification using unsupervised machine learning will be useful for identifying unknown spectra in new datasets.

\section{Conclusion}
We have shown that treating (energy-dependent) LEEM measurements as multi-dimensional datasets rather than as collection of images, opens rich opportunities for detailed and quantitative insights into complex material systems that go well beyond morphological and crystallographic characterization.

Three key steps are necessary to convert a stack of raw LEEM images into spectroscopic dataset with a greatly increased body of quantitative information.
First, we compensate for common detector artifacts such as camera dark count and non-uniform detector gain, which is crucial to quantitatively interpret LEEM images.
Second, by calibrating the channel plate gain and adjusting it during spectroscopic measurements, we can not only extend the dynamic range of the dataset by two orders of magnitude, but also convert image intensity into absolute reflectivity or electron intensity (provided the beam current is accurately measured). 
Third, we describe a drift-correction algorithm that is tailored for spectroscopic LEEM datasets where contrast inversions make many other approaches unfeasible. It relies on digital filtering and cross-correlation of every image to all other images and, without requiring large computation times, yields sub-pixel accuracy. It thus produces spectral LEEM data with high spatial resolution, i.e., true pixel-wise spectra.

This suite of techniques is already in regular use to obtain data from the ESCHER system in Leiden \cite{tromp2010new, tromp2013new, schramm2011low, jobst2016quantifying, DeJong2018-twist-angle, dejong2018intrinsic}.
In addition, the resulting spectral datasets enable more sophisticated data classification and visualization methods that rely on the spectrum (I(V)-curve) in every pixel. We demonstrate how we can use dimension reduction on the spectra to automatically compose images from only the six strongest spectroscopic features (PCA components). This approach produces rich color images that capture most of the features of the dataset and can thus give an intuitive view on complex material systems. 
Furthermore, we show that a relatively simple cluster analysis on those data sets of reduced dimensionality yields a quantitative representation of this information. Different materials within a field of view are automatically identified and statistical information such as the mean spectra and their spread per material can be extracted.

Treating LEEM measurements as multidimensional datasets as presented here will further strengthen the role of LEEM as a quantitative spectroscopic tool rather than as a pure imaging instrument, thus deepening its impact in the research and discovery of novel material systems. 
Furthermore, the presented techniques can be applied to related spectroscopic imaging techniques, such as energy-filtered PEEM~\cite{Renault2007XPEEM} or even adapted for use in scanning probe techniques such as scanning tunneling spectroscopy~\cite{Jesse2009PCA-STM,Yamanishi2018}.
To facilitate the use of the approaches discussed here 
, the test data as well as Python code is available online~\cite{..,..}.

\section*{Acknowledgements}
We thank Marcel Hesselberth and Douwe Scholma for their indispensable technical support and Christian Ott and Heiko Weber for the fabrication of the graphene on SiC samples. Funding: This work was supported by the Netherlands Organisation for Scientific Research (NWO/OCW) via the VENI Grant No. 680-47-447 (J.J.); and as part of the Frontiers of Nanoscience program.

\section*{References}
\bibliography{data-analysis-from-mendeley}

\begin{thebibliography}{10}
\expandafter\ifx\csname url\endcsname\relax
  \def\url#1{\texttt{#1}}\fi
\expandafter\ifx\csname urlprefix\endcsname\relax\def\urlprefix{URL }\fi
\expandafter\ifx\csname href\endcsname\relax
  \def\href#1#2{#2} \def\path#1{#1}\fi

\bibitem{Bauer1989}
E.~Bauer, M.~Mundschau, W.~Swiech, W.~Telieps,
  \href{https://www.sciencedirect.com/science/article/pii/0304399189900338
  https://linkinghub.elsevier.com/retrieve/pii/0304399189900338}{{Surface
  studies by low-energy electron microscopy (LEEM) and conventional UV
  photoemission electron microscopy (PEEM)}}, Ultramicroscopy 31~(1) (1989)
  49--57.
\newline\urlprefix\url{https://www.sciencedirect.com/science/article/pii/0304399189900338
  https://linkinghub.elsevier.com/retrieve/pii/0304399189900338}

\bibitem{dejong2018intrinsic}
T.~A. de~Jong, E.~E. Krasovskii, C.~Ott, R.~M. Tromp, S.~J. van~der Molen,
  J.~Jobst,
  \href{https://link.aps.org/doi/10.1103/PhysRevMaterials.2.104005}{{Intrinsic
  stacking domains in graphene on silicon carbide: A pathway for
  intercalation}}, Physical Review Materials 2~(10) (2018) 104005.
\newblock \href {https://doi.org/10.1103/PhysRevMaterials.2.104005}
  {\path{doi:10.1103/PhysRevMaterials.2.104005}}.
\newline\urlprefix\url{https://link.aps.org/doi/10.1103/PhysRevMaterials.2.104005}

\bibitem{flege2014intensity}
J.~I. Flege, E.~E. Krasovskii,
  \href{http://doi.wiley.com/10.1002/pssr.201409102}{{Intensity-voltage
  low-energy electron microscopy for functional materials characterization}},
  Physica Status Solidi - Rapid Research Letters 8~(6) (2014) 463--477.
\newblock \href {https://doi.org/10.1002/pssr.201409102}
  {\path{doi:10.1002/pssr.201409102}}.
\newline\urlprefix\url{http://doi.wiley.com/10.1002/pssr.201409102}

\bibitem{jobst2015nanoscale}
J.~Jobst, J.~Kautz, D.~Geelen, R.~M. Tromp, S.~J. van~der Molen,
  \href{http://www.nature.com/articles/ncomms9926}{{Nanoscale measurements of
  unoccupied band dispersion in few-layer graphene}}, Nature Communications
  6~(1) (2015) 8926.
\newblock \href {https://doi.org/10.1038/ncomms9926}
  {\path{doi:10.1038/ncomms9926}}.
\newline\urlprefix\url{http://www.nature.com/articles/ncomms9926}

\bibitem{jobst2016quantifying}
J.~Jobst, A.~J.~H. van~der Torren, E.~E. Krasovskii, J.~Balgley, C.~R. Dean,
  R.~M. Tromp, S.~J. van~der Molen,
  \href{http://www.nature.com/doifinder/10.1038/ncomms13621
  http://www.nature.com/articles/ncomms13621}{{Quantifying electronic band
  interactions in van der Waals materials using angle-resolved
  reflected-electron spectroscopy}}, Nature Communications 7~(1) (2016) 13621.
\newblock \href {https://doi.org/10.1038/ncomms13621}
  {\path{doi:10.1038/ncomms13621}}.
\newline\urlprefix\url{http://www.nature.com/doifinder/10.1038/ncomms13621
  http://www.nature.com/articles/ncomms13621}

\bibitem{Schmid1995}
A.~Schmid, W.~{\'{S}}wiȩch, C.~Rastomjee, B.~Rausenberger, W.~Engel,
  E.~Zeitler, A.~Bradshaw,
  \href{https://linkinghub.elsevier.com/retrieve/pii/003960289500128X}{{The
  chemistry of reaction-diffusion fronts investigated by microscopic LEED I–V
  fingerprinting}}, Surface Science 331-333~(PART A) (1995) 225--230.
\newline\urlprefix\url{https://linkinghub.elsevier.com/retrieve/pii/003960289500128X}

\bibitem{Frank2017}
B.~Frank, P.~Kahl, D.~Podbiel, G.~Spektor, M.~Orenstein, L.~Fu, T.~Weiss,
  M.~{Horn-von Hoegen}, T.~J. Davis, F.-J. {Meyer zu Heringdorf}, H.~Giessen,
  \href{http://advances.sciencemag.org/lookup/doi/10.1126/sciadv.1700721}{{Short-range
  surface plasmonics: Localized electron emission dynamics from a 60-nm spot on
  an atomically flat single-crystalline gold surface}}, Science Advances 3~(7)
  (2017) e1700721.
\newblock \href {https://doi.org/10.1126/sciadv.1700721}
  {\path{doi:10.1126/sciadv.1700721}}.
\newline\urlprefix\url{http://advances.sciencemag.org/lookup/doi/10.1126/sciadv.1700721}

\bibitem{locatelli2010corrugation}
A.~Locatelli, K.~R. Knox, D.~Cvetko, T.~O. Menteş, M.~A. Ni{\~{n}}o, S.~Wang,
  M.~B. Yilmaz, P.~Kim, R.~M. Osgood, A.~Morgante,
  \href{http://pubs.acs.org/doi/10.1021/nn101116n}{{Corrugation in Exfoliated
  Graphene: An Electron Microscopy and Diffraction Study}}, ACS Nano 4~(8)
  (2010) 4879--4889.
\newblock \href {https://doi.org/10.1021/nn101116n}
  {\path{doi:10.1021/nn101116n}}.
\newline\urlprefix\url{http://pubs.acs.org/doi/10.1021/nn101116n}

\bibitem{tromp2010new}
R.~Tromp, J.~Hannon, A.~Ellis, W.~Wan, A.~Berghaus, O.~Schaff,
  \href{https://linkinghub.elsevier.com/retrieve/pii/S0304399110000835}{{A new
  aberration-corrected, energy-filtered LEEM/PEEM instrument. I. Principles and
  design}}, Ultramicroscopy 110~(7) (2010) 852--861.
\newblock \href {https://doi.org/10.1016/j.ultramic.2010.03.005}
  {\path{doi:10.1016/j.ultramic.2010.03.005}}.
\newline\urlprefix\url{https://linkinghub.elsevier.com/retrieve/pii/S0304399110000835}

\bibitem{schramm2011low}
S.~Schramm, J.~Kautz, A.~Berghaus, O.~Schaff, R.~M. Tromp, S.~J. van~der Molen,
  Low-energy electron microscopy and spectroscopy with escher: Status and
  prospects, IBM Journal of Research and Development 55~(4) (2011) 1--1.
\newblock \href {https://doi.org/10.1147/JRD.2011.2150691}
  {\path{doi:10.1147/JRD.2011.2150691}}.

\bibitem{tromp2013new}
R.~M. Tromp, J.~B. Hannon, W.~Wan, A.~Berghaus, O.~Schaff,
  \href{https://www.sciencedirect.com/science/article/pii/S0304399112001866}{{A
  new aberration-corrected, energy-filtered LEEM/PEEM instrument II. Operation
  and results}}, Ultramicroscopy 127 (2013) 25--39.
\newblock \href {https://doi.org/10.1016/j.ultramic.2012.07.016}
  {\path{doi:10.1016/j.ultramic.2012.07.016}}.
\newline\urlprefix\url{https://www.sciencedirect.com/science/article/pii/S0304399112001866}

\bibitem{emtsev2009towards}
J.~R{\"{o}}hrl, E.~Rotenberg, J.~Jobst, T.~Seyller, D.~Waldmann, L.~Ley, J.~L.
  McChesney, G.~L. Kellogg, A.~K. Schmid, K.~Horn, S.~A. Reshanov, A.~Bostwick,
  T.~Ohta, H.~B. Weber, K.~V. Emtsev,
  \href{http://dx.doi.org/10.1038/nmat2382}{{Towards wafer-size graphene layers
  by atmospheric pressure graphitization of silicon carbide}}, Nature Materials
  8~(3) (2009) 203--207.
\newblock \href {https://doi.org/10.1038/nmat2382}
  {\path{doi:10.1038/nmat2382}}.
\newline\urlprefix\url{http://dx.doi.org/10.1038/nmat2382}

\bibitem{speck-QFMLG}
F.~Speck, J.~Jobst, F.~Fromm, M.~Ostler, D.~Waldmann, M.~Hundhausen, H.~B.
  Weber, T.~Seyller,
  \href{http://link.aip.org/link/APPLAB/v99/i12/p122106/s1{\&}Agg=doi
  http://aip.scitation.org/doi/10.1063/1.3643034}{{The quasi-free-standing
  nature of graphene on H-saturated SiC(0001)}}, Applied Physics Letters
  99~(12) (2011) 122106.
\newblock \href {https://doi.org/10.1063/1.3643034}
  {\path{doi:10.1063/1.3643034}}.
\newline\urlprefix\url{http://link.aip.org/link/APPLAB/v99/i12/p122106/s1{\&}Agg=doi
  http://aip.scitation.org/doi/10.1063/1.3643034}

\bibitem{riedl-QFMLG}
C.~Riedl, C.~Coletti, T.~Iwasaki, A.~A. Zakharov, U.~Starke,
  \href{http://link.aps.org/doi/10.1103/PhysRevLett.103.246804}{{Quasi-Free-Standing
  Epitaxial Graphene on SiC Obtained by Hydrogen Intercalation}}, Physical
  Review Letters 103~(24) (2009) 246804.
\newblock \href {https://doi.org/10.1103/PhysRevLett.103.246804}
  {\path{doi:10.1103/PhysRevLett.103.246804}}.
\newline\urlprefix\url{http://link.aps.org/doi/10.1103/PhysRevLett.103.246804}

\bibitem{hibino2008microscopic}
H.~Hibino, H.~Kageshima, F.~Maeda, M.~Nagase, Y.~Kobayashi, H.~Yamaguchi,
  \href{https://link.aps.org/doi/10.1103/PhysRevB.77.075413}{{Microscopic
  thickness determination of thin graphite films formed on SiC from quantized
  oscillation in reflectivity of low-energy electrons}}, Physical Review B
  77~(7) (2008) 075413.
\newblock \href {https://doi.org/10.1103/PhysRevB.77.075413}
  {\path{doi:10.1103/PhysRevB.77.075413}}.
\newline\urlprefix\url{https://link.aps.org/doi/10.1103/PhysRevB.77.075413}

\bibitem{feenstra2013low}
R.~M. Feenstra, N.~Srivastava, Q.~Gao, M.~Widom, B.~Diaconescu, T.~Ohta, G.~L.
  Kellogg, J.~T. Robinson, I.~V. Vlassiouk,
  \href{https://link.aps.org/doi/10.1103/PhysRevB.87.041406}{{Low-energy
  electron reflectivity from graphene}}, Physical Review B 87~(4) (2013)
  041406.
\newblock \href {https://doi.org/10.1103/PhysRevB.87.041406}
  {\path{doi:10.1103/PhysRevB.87.041406}}.
\newline\urlprefix\url{https://link.aps.org/doi/10.1103/PhysRevB.87.041406}

\bibitem{hibino2009stacking}
H.~Hibino, S.~Mizuno, H.~Kageshima, M.~Nagase, H.~Yamaguchi,
  \href{https://link.aps.org/doi/10.1103/PhysRevB.80.085406}{{Stacking domains
  of epitaxial few-layer graphene on SiC(0001)}}, Physical Review B 80~(8)
  (2009) 085406.
\newblock \href {https://doi.org/10.1103/PhysRevB.80.085406}
  {\path{doi:10.1103/PhysRevB.80.085406}}.
\newline\urlprefix\url{https://link.aps.org/doi/10.1103/PhysRevB.80.085406}

\bibitem{dejong2019data}
{de Jong, T.A. (Tobias)}, {Jobst, J. (Johannes)},
  \href{https://data.4tu.nl/repository/uuid:7f672638-66f6-4ec3-a16c-34181cc45202}{Data
  underlying the paper: Quantitative analysis of spectroscopic low energy
  electron microscopy data} (2019).
\newblock \href
  {https://doi.org/10.4121/uuid:7f672638-66f6-4ec3-a16c-34181cc45202}
  {\path{doi:10.4121/uuid:7f672638-66f6-4ec3-a16c-34181cc45202}}.
\newline\urlprefix\url{https://data.4tu.nl/repository/uuid:7f672638-66f6-4ec3-a16c-34181cc45202}

\bibitem{dejong2018intrinsicdata}
T.~A. {De Jong}, E.~E. Krasovskii, C.~Ott, R.~M. Tromp, S.~J. {Van Der Molen},
  J.~Jobst,
  \href{https://data.4tu.nl/repository/uuid:a7ff07f4-0ac8-4778-bec9-636532cfcfc1}{{Data
  underlying the paper: Intrinsic Stacking domains in graphene on silicon
  carbide: a pathway for intercalation}} (2018).
\newblock \href
  {https://doi.org/10.4121/UUID:A7FF07F4-0AC8-4778-BEC9-636532CFCFC1}
  {\path{doi:10.4121/UUID:A7FF07F4-0AC8-4778-BEC9-636532CFCFC1}}.
\newline\urlprefix\url{https://data.4tu.nl/repository/uuid:a7ff07f4-0ac8-4778-bec9-636532cfcfc1}

\bibitem{VanGastel2009Medipix}
R.~van Gastel, I.~Sikharulidze, S.~Schramm, J.~Abrahams, B.~Poelsema, R.~Tromp,
  S.~van~der Molen,
  \href{https://www.sciencedirect.com/science/article/pii/S0304399109001983}{{Medipix
  2 detector applied to low energy electron microscopy}}, Ultramicroscopy
  110~(1) (2009) 33--35.
\newblock \href {https://doi.org/10.1016/J.ULTRAMIC.2009.09.002}
  {\path{doi:10.1016/J.ULTRAMIC.2009.09.002}}.
\newline\urlprefix\url{https://www.sciencedirect.com/science/article/pii/S0304399109001983}

\bibitem{widenhorn2002temperature}
R.~Widenhorn, M.~M. Blouke, A.~Weber, A.~Rest, E.~Bodegom,
  \href{http://proceedings.spiedigitallibrary.org/proceeding.aspx?articleid=876982}{{Temperature
  dependence of dark current in a CCD}}, Vol. 4669, International Society for
  Optics and Photonics, 2002, pp. 193--201.
\newblock \href {https://doi.org/10.1117/12.463446}
  {\path{doi:10.1117/12.463446}}.
\newline\urlprefix\url{http://proceedings.spiedigitallibrary.org/proceeding.aspx?articleid=876982}

\bibitem{widenhorn2010exposure}
R.~Widenhorn, J.~C. Dunlap, E.~Bodegom,
  \href{http://ieeexplore.ieee.org/document/5378633/}{{Exposure Time Dependence
  of Dark Current in CCD Imagers}}, IEEE Transactions on Electron Devices
  57~(3) (2010) 581--587.
\newblock \href {https://doi.org/10.1109/TED.2009.2038649}
  {\path{doi:10.1109/TED.2009.2038649}}.
\newline\urlprefix\url{http://ieeexplore.ieee.org/document/5378633/}

\bibitem{widenhorn2001meyer}
R.~Widenhorn, L.~M{\"{u}}ndermann, A.~Rest, E.~Bodegom, {Meyer-Neldel rule for
  dark current in charge-coupled devices}, Journal of Applied Physics 89~(12)
  (2001) 8179--8182.
\newblock \href {https://doi.org/10.1063/1.1372365}
  {\path{doi:10.1063/1.1372365}}.

\bibitem{seibert1998flat}
J.~A. Seibert, J.~M. Boone, K.~K. Lindfors,
  \href{http://proceedings.spiedigitallibrary.org/proceeding.aspx?doi=10.1117/12.317034}{{Flat-field
  correction technique for digital detectors}}, Vol. 3336, International
  Society for Optics and Photonics, 1998, p. 348.
\newblock \href {https://doi.org/10.1117/12.317034}
  {\path{doi:10.1117/12.317034}}.
\newline\urlprefix\url{http://proceedings.spiedigitallibrary.org/proceeding.aspx?doi=10.1117/12.317034}

\bibitem{yu2010phase}
R.~Yu, S.~Kennedy, D.~Paganin, D.~Jesson,
  \href{https://linkinghub.elsevier.com/retrieve/pii/S096843280900170X}{{Phase
  retrieval low energy electron microscopy}}, Micron 41~(3) (2010) 232--238.
\newblock \href {https://doi.org/10.1016/j.micron.2009.10.010}
  {\path{doi:10.1016/j.micron.2009.10.010}}.
\newline\urlprefix\url{https://linkinghub.elsevier.com/retrieve/pii/S096843280900170X}

\bibitem{Kennedy-MM-distortions}
S.~M. Kennedy, C.~X. Zheng, W.~X. Tang, D.~M. Paganin, D.~E. Jesson,
  \href{http://rspa.royalsocietypublishing.org/cgi/doi/10.1098/rspa.2011.0204
  http://rspa.royalsocietypublishing.org/cgi/doi/10.1098/rspa.2010.0093}{{Laplacian
  image contrast in mirror electron microscopy}}, Proceedings of the Royal
  Society A: Mathematical, Physical and Engineering Sciences 466~(2124) (2010)
  2857--2874.
\newblock \href {https://doi.org/10.1098/rspa.2010.0093}
  {\path{doi:10.1098/rspa.2010.0093}}.
\newline\urlprefix\url{http://rspa.royalsocietypublishing.org/cgi/doi/10.1098/rspa.2011.0204
  http://rspa.royalsocietypublishing.org/cgi/doi/10.1098/rspa.2010.0093}

\bibitem{jobst2018quantifying}
J.~Jobst, L.~M. Boers, C.~Yin, J.~Aarts, R.~M. Tromp, S.~J. van~der Molen,
  \href{https://linkinghub.elsevier.com/retrieve/pii/S0304399118304182}{{Quantifying
  work function differences using low-energy electron microscopy: The case of
  mixed-terminated strontium titanate}}, Ultramicroscopy 200 (2019) 43--49.
\newblock \href {https://doi.org/10.1016/j.ultramic.2019.02.018}
  {\path{doi:10.1016/j.ultramic.2019.02.018}}.
\newline\urlprefix\url{https://linkinghub.elsevier.com/retrieve/pii/S0304399118304182}

\bibitem{Schramm2012Intrinsic}
S.~M. Schramm, S.~J. van~der Molen, R.~M. Tromp,
  \href{https://link.aps.org/doi/10.1103/PhysRevLett.109.163901}{{Intrinsic
  Instability of Aberration-Corrected Electron Microscopes}}, Physical Review
  Letters 109~(16) (2012) 163901.
\newblock \href {https://doi.org/10.1103/PhysRevLett.109.163901}
  {\path{doi:10.1103/PhysRevLett.109.163901}}.
\newline\urlprefix\url{https://link.aps.org/doi/10.1103/PhysRevLett.109.163901}

\bibitem{hamamatsuMCP}
Hamamatsu Photonics K.K.,, accessed on July 9, 2019.
\newblock
  \href{https://www.hamamatsu.com/resources/pdf/etd/MCP_TMCP0002E.pdf}{[link]}.
\newline\urlprefix\url{https://www.hamamatsu.com/resources/pdf/etd/MCP_TMCP0002E.pdf}

\bibitem{Foroosh2002}
H.~Foroosh, J.~Zerubia, M.~Berthod,
  \href{http://ieeexplore.ieee.org/document/988953/}{{Extension of phase
  correlation to subpixel registration}}, IEEE Transactions on Image Processing
  11~(3) (2002) 188--200.
\newblock \href {https://doi.org/10.1109/83.988953}
  {\path{doi:10.1109/83.988953}}.
\newline\urlprefix\url{http://ieeexplore.ieee.org/document/988953/}

\bibitem{Klein2010elastix}
S.~Klein, M.~Staring, K.~Murphy, M.~Viergever, J.~Pluim,
  \href{http://ieeexplore.ieee.org/document/5338015/}{{elastix: A Toolbox for
  Intensity-Based Medical Image Registration}}, IEEE Transactions on Medical
  Imaging 29~(1) (2010) 196--205.
\newblock \href {https://doi.org/10.1109/TMI.2009.2035616}
  {\path{doi:10.1109/TMI.2009.2035616}}.
\newline\urlprefix\url{http://ieeexplore.ieee.org/document/5338015/}

\bibitem{shamonin2014fast}
D.~Shamonin, E.~E. Bron, B.~P. Lelieveldt, M.~Smits, S.~Klein, M.~Staring,
  \href{http://journal.frontiersin.org/article/10.3389/fninf.2013.00050/abstract}{{Fast
  parallel image registration on CPU and GPU for diagnostic classification of
  Alzheimer's disease}}, Frontiers in Neuroinformatics 7 (2013) 50.
\newblock \href {https://doi.org/10.3389/fninf.2013.00050}
  {\path{doi:10.3389/fninf.2013.00050}}.
\newline\urlprefix\url{http://journal.frontiersin.org/article/10.3389/fninf.2013.00050/abstract}

\bibitem{IsmailiAalaoui2008}
E.~M. {Ismaili Aalaoui}, E.~Ibn-Elhaj,
  \href{http://www.hindawi.com/journals/jece/2008/417915/}{{Estimation of
  Subpixel Motion Using Bispectrum}}, Research Letters in Signal Processing
  2008~(13) (2008) 1--5.
\newblock \href {https://doi.org/10.1155/2008/417915}
  {\path{doi:10.1155/2008/417915}}.
\newline\urlprefix\url{http://www.hindawi.com/journals/jece/2008/417915/}

\bibitem{guizar2008efficient}
M.~Guizar-Sicairos, S.~T. Thurman, J.~R. Fienup,
  \href{https://www.osapublishing.org/abstract.cfm?URI=ol-33-2-156}{{Efficient
  subpixel image registration algorithms}}, Optics Letters 33~(2) (2008) 156.
\newblock \href {https://doi.org/10.1364/OL.33.000156}
  {\path{doi:10.1364/OL.33.000156}}.
\newline\urlprefix\url{https://www.osapublishing.org/abstract.cfm?URI=ol-33-2-156}

\bibitem{Tzimiropoulos2011}
G.~Tzimiropoulos, V.~Argyriou, T.~Stathaki,
  \href{http://ieeexplore.ieee.org/document/5648352/}{{Subpixel Registration
  With Gradient Correlation}}, IEEE Transactions on Image Processing 20~(6)
  (2011) 1761--1767.
\newblock \href {https://doi.org/10.1109/TIP.2010.2095867}
  {\path{doi:10.1109/TIP.2010.2095867}}.
\newline\urlprefix\url{http://ieeexplore.ieee.org/document/5648352/}

\bibitem{Schaffer2004Automated}
B.~Schaffer, W.~Grogger, G.~Kothleitner,
  \href{https://www.sciencedirect.com/science/article/pii/S0304399104001652
  http://linkinghub.elsevier.com/retrieve/pii/S0304399104001652}{{Automated
  spatial drift correction for EFTEM image series}}, Ultramicroscopy 102~(1)
  (2004) 27--36.
\newblock \href {https://doi.org/10.1016/j.ultramic.2004.08.003}
  {\path{doi:10.1016/j.ultramic.2004.08.003}}.
\newline\urlprefix\url{https://www.sciencedirect.com/science/article/pii/S0304399104001652
  http://linkinghub.elsevier.com/retrieve/pii/S0304399104001652}

\bibitem{tobias2019github}
{de Jong, T.A.}, \href{https://github.com/TAdeJong/LEEM-analysis}{Quantitative
  data analysis for spectroscopic {LEEM}} (2019).
\newline\urlprefix\url{https://github.com/TAdeJong/LEEM-analysis}

\bibitem{adelson1984pyramid}
E.~Adelson, P.~Burt, C.~Anderson, J.~Ogden, J.~Bergen, Pyramid methods in image
  processing, RCA engineer 29~(6) (1984) 33--41.

\bibitem{thevenaz1998pyramid}
P.~Thevenaz, U.~Ruttimann, M.~Unser,
  \href{http://ieeexplore.ieee.org/document/650848/}{{A pyramid approach to
  subpixel registration based on intensity}}, IEEE Transactions on Image
  Processing 7~(1) (1998) 27--41.
\newblock \href {https://doi.org/10.1109/83.650848}
  {\path{doi:10.1109/83.650848}}.
\newline\urlprefix\url{http://ieeexplore.ieee.org/document/650848/}

\bibitem{wold1987principal}
S.~Wold, K.~Esbensen, P.~Geladi,
  \href{https://www.sciencedirect.com/science/article/pii/0169743987800849
  https://linkinghub.elsevier.com/retrieve/pii/0169743987800849}{{Principal
  component analysis}}, Chemometrics and Intelligent Laboratory Systems 2~(1-3)
  (1987) 37--52.
\newblock \href {https://doi.org/10.1016/0169-7439(87)80084-9}
  {\path{doi:10.1016/0169-7439(87)80084-9}}.
\newline\urlprefix\url{https://www.sciencedirect.com/science/article/pii/0169743987800849
  https://linkinghub.elsevier.com/retrieve/pii/0169743987800849}

\bibitem{Jesse2009PCA-STM}
S.~Jesse, S.~V. Kalinin,
  \href{http://stacks.iop.org/0957-4484/20/i=8/a=085714?key=crossref.c67bebb898c90cc5d61d15749a4d5c63}{{Principal
  component and spatial correlation analysis of spectroscopic-imaging data in
  scanning probe microscopy}}, Nanotechnology 20~(8) (2009) 085714.
\newblock \href {https://doi.org/10.1088/0957-4484/20/8/085714}
  {\path{doi:10.1088/0957-4484/20/8/085714}}.
\newline\urlprefix\url{http://stacks.iop.org/0957-4484/20/i=8/a=085714?key=crossref.c67bebb898c90cc5d61d15749a4d5c63}

\bibitem{abdi2010principal}
H.~Abdi, L.~J. Williams,
  \href{http://doi.wiley.com/10.1002/wics.101}{{Principal component analysis}},
  Wiley Interdisciplinary Reviews: Computational Statistics 2~(4) (2010)
  433--459.
\newblock \href {https://doi.org/10.1002/wics.101}
  {\path{doi:10.1002/wics.101}}.
\newline\urlprefix\url{http://doi.wiley.com/10.1002/wics.101}

\bibitem{halko2011finding}
N.~Halko, P.~G. Martinsson, J.~A. Tropp,
  \href{http://epubs.siam.org/doi/10.1137/090771806
  http://arxiv.org/abs/0909.4061}{{Finding Structure with Randomness:
  Probabilistic Algorithms for Constructing Approximate Matrix
  Decompositions}}, SIAM Review 53~(2) (2011) 217--288.
\newblock \href {http://arxiv.org/abs/0909.4061} {\path{arXiv:0909.4061}},
  \href {https://doi.org/10.1137/090771806} {\path{doi:10.1137/090771806}}.
\newline\urlprefix\url{http://epubs.siam.org/doi/10.1137/090771806
  http://arxiv.org/abs/0909.4061}

\bibitem{kovesi2015good}
P.~Kovesi, \href{http://arxiv.org/abs/1509.03700}{{Good Colour Maps: How to
  Design Them}}, arXiv preprint arXiv:1509.03700 (2015).
\newblock \href {http://arxiv.org/abs/1509.03700} {\path{arXiv:1509.03700}}.
\newline\urlprefix\url{http://arxiv.org/abs/1509.03700}

\bibitem{lloyd1982least}
S.~Lloyd, \href{http://ieeexplore.ieee.org/document/1056489/}{{Least squares
  quantization in PCM}}, IEEE Transactions on Information Theory 28~(2) (1982)
  129--137.
\newblock \href {https://doi.org/10.1109/TIT.1982.1056489}
  {\path{doi:10.1109/TIT.1982.1056489}}.
\newline\urlprefix\url{http://ieeexplore.ieee.org/document/1056489/}

\bibitem{DeJong2018-twist-angle}
T.~A. de~Jong, J.~Jobst, H.~Yoo, E.~E. Krasovskii, P.~Kim, S.~J. van~der Molen,
  \href{http://arxiv.org/abs/1806.05155
  http://doi.wiley.com/10.1002/pssb.201800191}{{Measuring the Local Twist Angle
  and Layer Arrangement in Van der Waals Heterostructures}}, physica status
  solidi (b) (2018) 1800191\href {https://doi.org/10.1002/pssb.201800191}
  {\path{doi:10.1002/pssb.201800191}}.
\newline\urlprefix\url{http://arxiv.org/abs/1806.05155
  http://doi.wiley.com/10.1002/pssb.201800191}

\bibitem{Renault2007XPEEM}
O.~Renault, N.~Barrett, A.~Bailly, L.~Zagonel, D.~Mariolle, J.~Cezar,
  N.~Brookes, K.~Winkler, B.~Kr{\"{o}}mker, D.~Funnemann,
  \href{https://www.sciencedirect.com/science/article/pii/S0039602807006085}{{Energy-filtered
  XPEEM with NanoESCA using synchrotron and laboratory X-ray sources:
  Principles and first demonstrated results}}, Surface Science 601~(20) (2007)
  4727--4732.
\newblock \href {https://doi.org/10.1016/j.susc.2007.05.061}
  {\path{doi:10.1016/j.susc.2007.05.061}}.
\newline\urlprefix\url{https://www.sciencedirect.com/science/article/pii/S0039602807006085}

\bibitem{Yamanishi2018}
J.~Yamanishi, S.~Iwase, N.~Ishida, D.~Fujita,
  \href{https://www.sciencedirect.com/science/article/pii/S0169433217327721}{{Multivariate
  analysis for scanning tunneling spectroscopy data}}, Applied Surface Science
  428 (2018) 186--190.
\newblock \href {https://doi.org/10.1016/J.APSUSC.2017.09.124}
  {\path{doi:10.1016/J.APSUSC.2017.09.124}}.
\newline\urlprefix\url{https://www.sciencedirect.com/science/article/pii/S0169433217327721}

\bibitem{dask}
{Dask Development Team}, \href{https://dask.org}{{Dask: Library for dynamic
  task scheduling}} (2016).
\newline\urlprefix\url{https://dask.org}

\bibitem{matthew_rocklin-proc-scipy-2015}
M.~Rocklin,
  \href{https://conference.scipy.org/proceedings/scipy2015/matthew{\_}rocklin.html}{{Dask:
  Parallel Computation with Blocked algorithms and Task Scheduling}}, in:
  K.~Huff, J.~Bergstra (Eds.), Proceedings of the 14th Python in Science
  Conference, 2015, pp. 126--132.
\newblock \href {https://doi.org/10.25080/Majora-7b98e3ed-013}
  {\path{doi:10.25080/Majora-7b98e3ed-013}}.
\newline\urlprefix\url{https://conference.scipy.org/proceedings/scipy2015/matthew{\_}rocklin.html}

\bibitem{van2014scikit}
S.~van~der Walt, J.~L. Sch{\"{o}}nberger, J.~Nunez-Iglesias, F.~Boulogne, J.~D.
  Warner, N.~Yager, E.~Gouillart, T.~Yu,
  \href{https://peerj.com/articles/453}{{scikit-image: image processing in
  Python}}, PeerJ 2 (2014) e453.
\newblock \href {https://doi.org/10.7717/peerj.453}
  {\path{doi:10.7717/peerj.453}}.
\newline\urlprefix\url{https://peerj.com/articles/453}

\bibitem{scipylib}
E.~Jones, T.~Oliphant, P.~Peterson, E.~Al., \href{http://www.scipy.org}{{SciPy:
  Open source scientific tools for Python, http://www.scipy.org}} (2001).
\newline\urlprefix\url{http://www.scipy.org}

\bibitem{branch1999subspace}
M.~A. Branch, T.~F. Coleman, Y.~Li,
  \href{http://epubs.siam.org/doi/10.1137/S1064827595289108}{{A Subspace,
  Interior, and Conjugate Gradient Method for Large-Scale Bound-Constrained
  Minimization Problems}}, SIAM Journal on Scientific Computing 21~(1) (1999)
  1--23.
\newblock \href {https://doi.org/10.1137/S1064827595289108}
  {\path{doi:10.1137/S1064827595289108}}.
\newline\urlprefix\url{http://epubs.siam.org/doi/10.1137/S1064827595289108}

\bibitem{byrd1988approximate}
R.~H. Byrd, R.~B. Schnabel, G.~A. Shultz,
  \href{http://link.springer.com/10.1007/BF01580735}{{Approximate solution of
  the trust region problem by minimization over two-dimensional subspaces}},
  Mathematical Programming 40-40~(1-3) (1988) 247--263.
\newblock \href {https://doi.org/10.1007/BF01580735}
  {\path{doi:10.1007/BF01580735}}.
\newline\urlprefix\url{http://link.springer.com/10.1007/BF01580735}

\bibitem{perez2007ipython}
F.~Perez, B.~E. Granger, \href{http://ipython.org
  http://ieeexplore.ieee.org/document/4160251/}{{IPython: A System for
  Interactive Scientific Computing}}, Computing in Science {\&} Engineering
  9~(3) (2007) 21--29.
\newblock \href {https://doi.org/10.1109/MCSE.2007.53}
  {\path{doi:10.1109/MCSE.2007.53}}.
\newline\urlprefix\url{http://ipython.org
  http://ieeexplore.ieee.org/document/4160251/}

\bibitem{pedregosa2011scikit}
F.~Pedregosa, G.~Varoquaux, A.~Gramfort, V.~Michel, B.~Thirion, O.~Grisel,
  M.~Blondel, A.~M{\"{u}}ller, J.~Nothman, G.~Louppe, P.~Prettenhofer,
  R.~Weiss, V.~Dubourg, J.~Vanderplas, A.~Passos, D.~Cournapeau, M.~Brucher,
  M.~Perrot, {\'{E}}.~Duchesnay,
  \href{http://arxiv.org/abs/1201.0490}{{Scikit-learn: Machine Learning in
  Python}}, Journal of machine learning research 12~(Oct) (2012) 2825--2830.
\newblock \href {http://arxiv.org/abs/1201.0490} {\path{arXiv:1201.0490}}.
\newline\urlprefix\url{http://arxiv.org/abs/1201.0490}

\end{thebibliography}

\appendix
\FloatBarrier
\newpage
\section{Supplemental information}

\subsection{Implementation}
The implementation presented here is programmed in Python, making extensive use of \texttt{dask}~\cite{dask, matthew_rocklin-proc-scipy-2015}. 
This open source library, developed and maintained by Anaconda Inc., in particular its \texttt{Array} submodule, enables easy parallelization of array operations in common \texttt{numpy} syntax. 
It allows the lazy definition of computational operations on data, forming a task graph describing the computations to be performed.
Delaying actual computation until explicitly called for enables \texttt{dask} to easily parallelize and stream computations, efficiently using all cores of a single computer.
It should even allow to easily scale up to compute clusters.

The implementation uses of \texttt{dask} in combination with \texttt{scikit-image} code for blocked filtering of images~\cite{van2014scikit}.

The complete set of correlations is written in \texttt{dask} via the FFT functions and the multiplication in Fourier space. This, via \texttt{dask}'s task graph, allows every forward FFT to be computed only once for each image, as opposed to the $N$ times for the naive implementation. 

The optimization step uses \texttt{scipy} code, in particular the \texttt{least-squares} routine in the \texttt{optimization} module in combination with an explicit \texttt{scipy.sparse} Jacobian.~\cite{van2014scikit, scipylib, branch1999subspace, byrd1988approximate}.
We found that the use of an explicit Jacobian significantly reduced computation time and memory use for larger optimizations.

For easy interfacing with the user, \texttt{jupyter-notebook} and \texttt{ipython-widgets} are used~\cite{perez2007ipython}.
For the dimension reduction using principal component analysis and clustering \texttt{dask-ml} and \texttt{scikit-learn} are applied~\cite{pedregosa2011scikit}.

\subsection{Benchmark methodology}
Benchmarks were run on a desktop PC with an Intel \texttt{i7-7700K} running at 4.20 GHz with 32 GB of RAM and Windows 10 x64 installed. The full stack of images was read from separate \texttt{TIFF} files on a Toshiba XG5 NVMe SSD for each run during the filter and cross correlation phase and written to a uncompressed \texttt{zarr} archive on the same SSD during the shift and write phase.

The benchmark was run in a Jupyter notebook, with an installation of the Anaconda environment of Python 3.7 with at least the following packages installed:\\

\texttt{numpy=1.16.3, matplotlib=3.0.3, dask=1.2.0,}\\ 
    \texttt{distributed=1.27.1, scikit-image=0.15.0, scipy=1.2.1, jupyterlab, scikit-learn=0.20.3, dask-ml=0.12.0,}\\
    \texttt{xarray=0.12.0, h5py=2.9.0, mkl=2019.3, mkl\_fft=1.0.12, numba=0.43.1, zarr=2.2.0}
    
Accuracy benchmarks were performed with $512\times512$ images, time benchmarks on subimages of $256\times256$.

\begin{figure*}
\centering
\includegraphics[width=1.8\columnwidth]{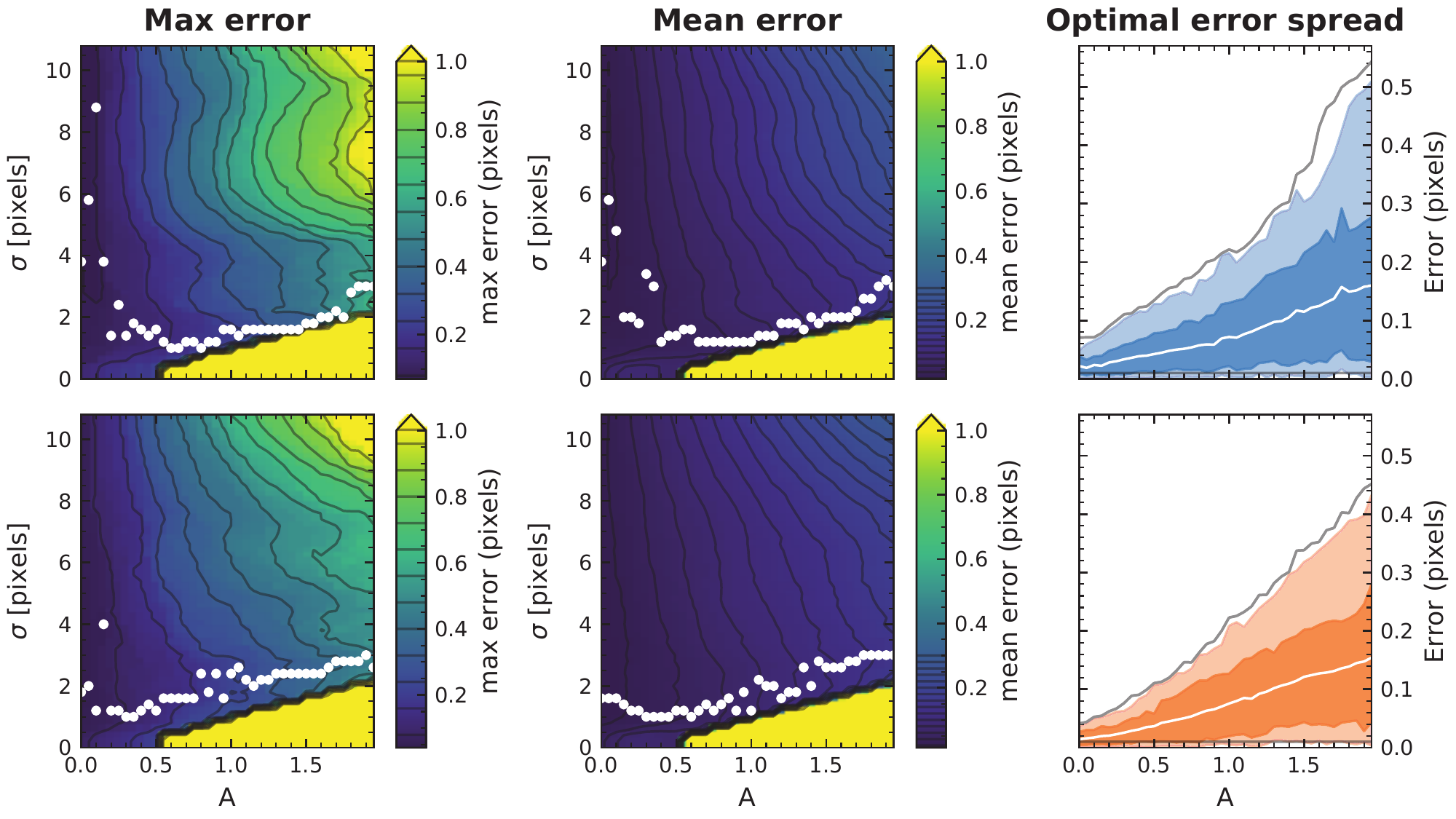}
\caption{Maximum and mean error in $x$ and $y$ shift as calculated by the $N^2$ algorithm, with a different random seed compared to Fig.\ \ref{fig:simulationresults}. The optimal value of the Gaussian smoothing $s$ as a function of added noise amplitude $A$ is drawn in white. Contour lines are added as a guide to the eye.}
\label{suppfig:simulationresults}
\end{figure*}
\begin{figure*}
\centering
\includegraphics[width=1.8\columnwidth]{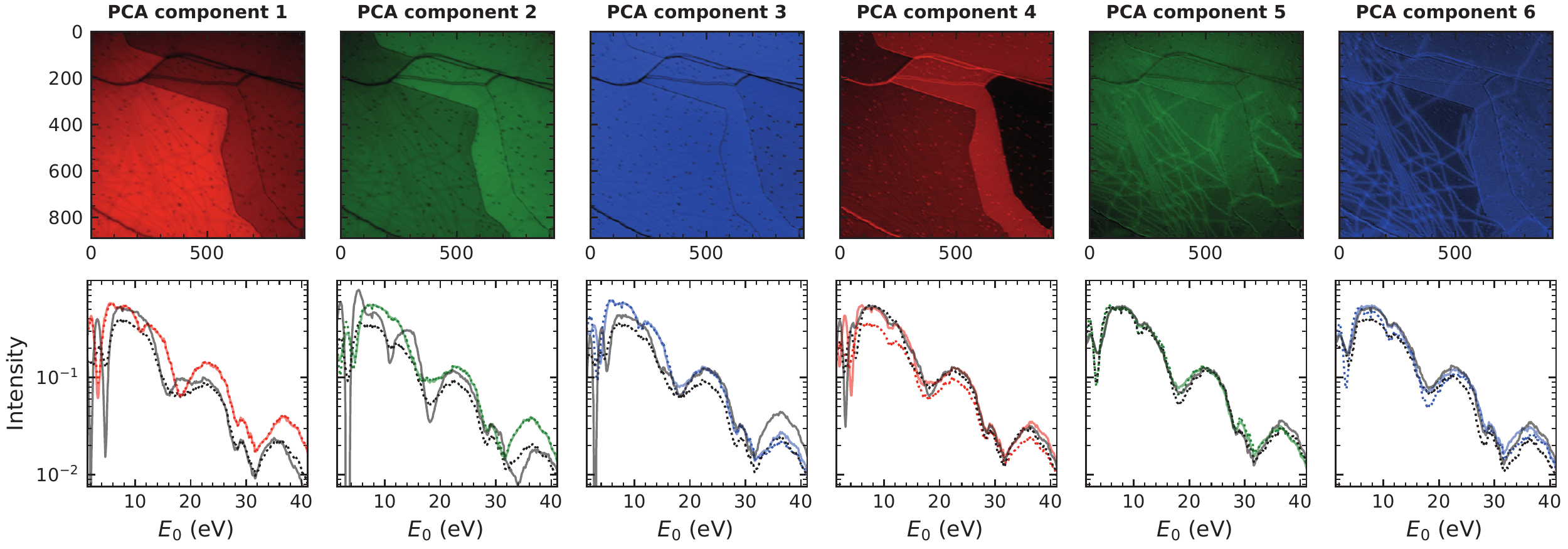}\\
\caption{\subf{a} Images of the first six PCA components for the Bright Field sample data set and the spectra corresponding to the maximum and minimum of this component occurring in the dataset.
}
\label{suppfig:PCAcomponents}
\end{figure*}
\end{document}